\documentclass[prd,aps,showpacs,groupedaddress,superscriptaddress,nofootinbib]{revtex4}
\usepackage{amssymb}
\usepackage{graphicx}
\usepackage{amsmath}
\usepackage{comment}

\newcommand{\be}{\begin{equation}}
\newcommand{\ee}{\end{equation}}
\newcommand{\bq}{\begin{eqnarray}}
\newcommand{\eq}{\end{eqnarray}}

\begin{document}
\title{The K-essence flow seen from the preferred frame $S_{V}$. A scalar field theory with Landau superfluid structure
}
\author{Rodrigo Francisco dos Santos}
\email{santosst1@gmail.com}
\affiliation{Department of Education-Belford Roxo City, Jd. Glaucia School Town, Pericles 15, pc 26195-140, Belford Roxo, RJ, Brazil}
\author{Luis Gustavo de Almeida}
\email{lgalmeida@gmail.com}
\affiliation{Federal University of Acre, Rio Branco, pc 69920-900, AC, Brazil}
\author{A. C. Amaro de Faria Jr}
\email{atoni.carlos@gmail.com}
\affiliation{Federal Technological University of Parana, Guarapuava, pc 85053-525, PR, Brazil}
\altaffiliation[]{Permanent address}
\affiliation{Advanced Studies Institute - IEAv, Sao Jose dos Campos, pc 12220-000, SP, Brazil}

\begin{abstract}

We study the hypothesis of deformation of the invariance of Lorentz transformations produced by the introduction of a universal minimum velocity relative to a preferred frame. Our goal with this job is  to apply this hypothesis to superfluids and study its consequences relating the minimum velocity to the idea of a fluid, with superfluid properties. In previous works we related the minimum velocity to the cosmological constant and even to cosmic inflation. Soon we could generate a hypothetical superfluid capable of modeling with characteristics of a cosmological fluid with dark energy properties.  The first excited state of this universal superfluid would be a preferred frame from which all other excited states are observed and then we would have a preferred frame $S_{V}$ associated with the critical Landau velocity, thus implying that the universal minimum velocity coincides with the critical Landau velocity, and the objects observed by the preferred frame are excited states of the superfluid. This coincidence between the concepts of minimum velocity and Landau's critical velocity makes Landau's critical velocity a type of limit velocity, modifying the usual causal structure of restricted relativity.  Formulating the phenomena in this preferred frame would have the advantage of providing a simple explanation for astrophysical and cosmological phenomena linked to a causal structure, which emerges from this construction and is very similar to causal structures linked to Gordon geometry and acoustic tachyons.
 We build a deformed relativistic Lagrangian, demonstrate its relation with a \emph{k}-essence Lagrangian and calculate the quantities associated with that Lagrangian. We also studied an irrotational fluid and verified the role of enthalpy associated with the minimum velocity structure.

\end{abstract}

\pacs{11.30.Qc}
\maketitle

\section{Introduction}

An important line of research in cosmology has been the model entitled \emph{k}-essence \cite{kan}.  The \emph{k}-essence models are scalar field models~\cite{k1,ka,k2,k3},  that alongside the quintessence models~\cite{q1,q2}, appear as alternatives to solve the problem of barotropic fluids with $w={p}/{\rho}<0$. Therefore, one of the most important guidelines in the study of \emph{k}-essence is the study of causal structures and their pathologies~\cite{kpato}. Recent developments \cite{kpato1} point to an approximation of the \emph{k}-essence hypothesis with superfluid models \cite{vortex,superverse}.  The line of gravitational superfluids models has received a lot of attention in recent years, in particular using  the framework of Carter~\cite{superrela,superrela1}, known as Carter's Multifluid Theory~\cite{supercarter}, and its relation of gravity~\cite{supercarter1,supercarter2}. This is very similar line to the line developed by Visser~\cite{v2010}. In this line, Bacetti \cite{bacejhep2} developed a treatment for dissipation relations deformed by a Lorentz symmetry breaking, which is very similar to the Landau formalism used in superfluids, also applicable to Planck scale dissipations \cite{volo22}. In 2017 Nassif~\emph{et al}~\cite{r1}, in an exotic but promising approach, starts the study of effects of  deformation of Lorentz Invariance introduced by a universal minimum velocity relative to a  preferred reference frame in an inflationary scenario~\cite{r2,r3} and gravitational collapse~\cite{gbec}. Nassif's works~\cite{N2007,N2008,N2013,N2015,N2016,N2019,N2022} have suggested a dispersion relation deformed by the invariant minimal speed, associated with a preferred frame~$S_V$. However, in 2022 Santos~\emph{et al}~\cite{nova} propose a totally new perspective for the deformation introduced in incipient Nassif's works, where they look for a hydrodynamic perspective for the minimal speed~\cite{rhidro,rhidro1}. This new interpretation aims to adjust the idea of a minimum velocity to a relativistic hydrodynamics scenario that approximates the gravitational interaction of phase transition phenomena~\cite{svt,svt2,qgv} metamaterials~\cite{meta,meta2} to a Relativistic Bose Einstein Condensate~\cite{rbec}.
That said, let's explicitly enumerate the objectives of our work 
\begin{enumerate}
    \item{The goal of this paper is to find a physical justification for the introduction of the minimum velocity\cite{N2007}, in this case the justification is the relationship of the Nassif minimum velocity to the Landau critical velocity. This coincidence is very promising, since superfluids have properties analogous to anti-gravity. We can therefore use them as analogues to the gravitational vacuum. }
    \item{Demonstrate that the introduction of the minimum velocity is equivalent to a violation of the Lorentz symmetry in an acoustic geometry}
    \item{To resume Nassif's uncertainty principle proposal\cite{N2013}, but with a description that makes it clear that the quantum characteristic of this treatment is analogous to that of a superfluid\cite{superverse}}
    \item{Demonstrate the Lorentz Invariance Violation in the same terms as Zloschatiev \cite{svt2}}
    \item{ To identify points of falsifiability of the existence of the preferred-frame $S_{V}$ \cite{N2007} we study the relation of the Einstein-Euller referential and the preferred-frame $S_{V}$. As we try to understand the relevance of the $S_{V}$ referential to other proposals of fluids that intend to model the dark energy\cite{r2}, in this paper the fluid chosen is \emph{k}-essence}
    \item{Performing the first movements to understand the energy mechanism, which would make each excited state appear in the hypothetical fluid, excited states that we can analogously compare with subatomic matter appearing in the gravitational vacuum on a cosmological scale could be a more complex Structures. Since we can hypothetically extrapolate the Jeans mechanism \cite{jeans}}
\end{enumerate}

This work is divided as follows:

In Section 2 we briefly review the \emph{k}-essence model and specify the kinetic \emph{k}-essence. 

In Section 3, we revisit the concepts associated with the introduction of minimum velocity in Lorentz transformations.

The Section 4 shows how the minimum speed corresponds to a critical speed in the Landau criterion, a relation to the Einstein-Euller referential, helping to clarify the role of the preferred-frame $S_{V}$. And we construct a emph{K}-Essence Lagrangian, taking into account the scalar product deformation, which introduces the reciprocal velocity

The Section 5 presents an approach where we seek to investigate some thermodynamic properties of an irrotational fluid considering the Shutz formalism~\cite{sch,sch2}. And rehearsing a discussion about the jumps of energy levels for the appearance of each excited state with the increment of the velocity $v$

In section 7 we present our conclusions and future perspectives.

\section{\emph{K}-essence  framework}
The \emph{k}-essence model is a theoretical framework used in hydrodynamics to describe the dynamics of a scalar field with a non-canonical kinetic term. This model introduces a new degree of freedom, which modifies the equation of state of the fluid and allows for the possibility of an accelerated expansion of the universe. The \emph{k}-essence model has been used to study a wide range of phenomena, including cosmic inflation~\cite{kNojiri2019,kOikonomou2021}, dark energy~\cite{kOdintsov2020,kOdintsov2021,kOikonomou2020,kOikonomou2021}, and the dynamics of cosmic structure formation.

We can define a \emph{k}-essence model~\cite{k1,ka} with the action 
\begin{equation}\label{first}
\mathcal{S}:=\int{d^{4}x}\sqrt{g}\mathcal{L}(\mathcal{X},\Phi). 
\end{equation}
The action~(\ref{first}) is associated to the lagrangian density~$\mathcal{L}(\mathcal{X},\Phi)$, where $\mathcal{X}$ is the scalar product defined as
\begin{equation}\label{sc}
\mathcal{X}:=g^{\mu\nu}\nabla_{\mu}\Phi\nabla_{\nu}\Phi.
\end{equation}
Here $\Phi$ is a scalar field and $\nabla$ is a covariant derivative operator linked to $g^{\mu\nu}$. It is very important notice that~$\mathcal{X}$ is a scalar product between two non-normalized four-vectors, with the presence of the normalized factor~$\nabla_{\mu}\Phi$. 
The conserved current associeted to~(\ref{first}) is obtained in~\cite{k2} 
\begin{equation}\label{cc}
\left(\mathcal{L}_{,\mathcal{X}}g^{\alpha\beta}-\mathcal{L}_{,\mathcal{XX}}\nabla^{\alpha}\Phi\nabla^{\beta}\Phi\right)\nabla_{\alpha}\nabla_{\beta}\Phi+\mathcal{L}_{,\Phi\mathcal{X}}g^{\alpha\beta}\nabla_{\alpha}\Phi\nabla_{\beta}\Phi=-\mathcal{L}_{\Phi},
\end{equation}
where $\mathcal{L}_{,\mathcal{X}}$ is the derivative for parameter $\mathcal{X}$, $\mathcal{L}_{,\mathcal{XX}}$ is second derivative with $\mathcal{X}$. Also, this $\mathcal{L}_{,\Phi}$ is the derivative with~$\Phi$ and~$\mathcal{L}_{,\Phi\mathcal{X}}$ is a second derivative. In the spatially homogeneous, case~$\mathcal{X}>0$, it reduces to
\begin{equation}\label{2}
\left(2\mathcal{X}\mathcal{L}_{,\mathcal{XX}}+\mathcal{L}_{,\mathcal{X}}\right)\ddot{\Phi}+\mathcal{L}_{,\mathcal{X}}(3H\dot{\Phi})+\mathcal{L}_{,\Phi\mathcal{X}}\dot{\Phi}^{2}-\mathcal{L}_{,\Phi}=0.    
\end{equation} 

To ensure solubility, we take $\left(2\mathcal{X}\mathcal{L}_{,\mathcal{XX}}+\mathcal{L}_{,\mathcal{X}}\right)\neq0$. We can further restrict this condition, calling it a condition of ``hyperbolicity'',
\begin{equation}\label{hiper}
    1+2\frac{\mathcal{X}\mathcal{L}_{,\mathcal{XX}}}{\mathcal{L}_{,\mathcal{X}}}\geq0,
\end{equation} thus imposing a condition on $\mathcal{L}_{,\mathcal{X}}$ and $\mathcal{L}_{,\mathcal{XX}}$.
\subsection{\emph{K}-essence purely kinetic: Effective Hydrodynamics Approach}
The lagrangean $\mathcal{L}(\mathcal{X},\Phi)$ eventuality showed a dependency only of~$\mathcal{X}$, so we can write 
\begin{equation}\label{puro}
\mathcal{L}(\mathcal{X},\Phi)\equiv\mathcal{L}(\mathcal{X}).
\end{equation} This is a particular case  called ``purelly kinetic \emph{k}-essence'' and imply in an isotropic and homogeneous preferred frame backgroud~\cite{k1,ka}. In this particular case we have $\mathcal{L}_{\Phi\mathcal{X}}=\mathcal{L}_{\Phi}=0$. The lagrangean~(\ref{first}) allow evaluation of momentum-energy tensor, $T_{\mu\nu}:=\frac{2}{\sqrt{g}}\frac{\delta\mathcal{L}}{\delta{g}_{\mu\nu}}$.  We also can write  the momentum-energy tensor as we did in~\cite{r2}:
\begin{equation}\label{tm}
T_{\mu\nu}=\mathcal{L}_{,\mathcal{X}}\nabla_{\mu}\Phi\nabla_{\nu}\Phi-\mathcal{L}(\mathcal{X})g_{\mu\nu}.
\end{equation}

The simple comparation of equation~(\ref{tm}) with energy-momentum tensor for a perfect fluid suggests the identification of several hydrodynamics variables. 
The pressure can be written as
 \begin{equation}\label{pressão}
p=\mathcal{L}(\mathcal{X}). 
 \end{equation}
 The kinetic term requires more work. We define the four-velocity
 \begin{equation}\label{nor4v}
 u_{\mu}:=\frac{\nabla_{\mu}\Phi}{\sqrt{2\mathcal{X}}},
\end{equation}
 replacing~(\ref{pressão}) and~(\ref{nor4v}) in~(\ref{tm}), we have density of energy
\begin{equation}\label{ro}
\rho=\mathcal{X}\mathcal{L}_{,\mathcal{X}}-\mathcal{L}(\mathcal{X}).
\end{equation}

We write now the energy-momentum tensor with \emph{k}-essence as source:
\begin{equation}\label{ktm}
T_{\mu\nu}=2\mathcal{X}\mathcal{L}_{,\mathcal{X}}u_{\mu}u_{\nu}-\mathcal{L}(\mathcal{X})g_{\mu\nu}.   
\end{equation}
For~(\ref{ktm}) exists a rest-frame~$u_{i}=0$ where the scalar field is locally isotropic.

Now we are going to discuss the energy conditions in a similar way of the work~\cite{k2}. The \emph{null energy condition}~(NEC), is respected
\begin{equation}\label{nec}
\mathcal{L}_{,\mathcal{X}}>0,
\end{equation} the \emph{weak energy condition}~(WEC), establishes 
\begin{equation}\label{wec}
    \mathcal{X}\mathcal{L}_{,\mathcal{X}}-\mathcal{L}(\mathcal{X})>0,
\end{equation}
and  the~(\ref{nec})  with~$\mathcal{L}>0$. Both conditions in force guarantee that we deal only with positive energy states. In the event of a violation, negative energy states arise.

The expression 
\begin{equation}\label{DEC}
    \frac{\mathcal{X}\mathcal{L}_{,\mathcal{X}}-\mathcal{L}(\mathcal{X})}{\mathcal{L}(\mathcal{X})}>0,
\end{equation}
is associated with the \emph{dominate energy condition}~(DEC), and~(\ref{DEC}) with~(\ref{nec}), generate the \emph{null dominant energy condition}~(NDEC). The violation of those conditions can introduce tachyons.

We define the barotropic parameter $w$ according to Visser et-al~\cite{v2010}, and establish criteria for a barotropic fluid:
\begin{itemize}
    \item Any zero-temperature fluid is automatically barotropic;
    \item Any non-zero temperature but isothermal fluid is automatically barotropic;
    \item Any zero-entropy fluid (superfluid) is automatically barotropic;
    \item Any isentropic fluid is automatically barotropic.
\end{itemize}

We write, then 
\begin{equation}\label{barro}
    w:=\frac{p}{\rho}=\frac{\mathcal{L}(\mathcal{X})}{\mathcal{X}\mathcal{L}_{,\mathcal{X}}-\mathcal{L}},
\end{equation}
and get the sound propagation, according to~\cite{k3}
\begin{equation}\label{som}
  c^{2}_{s}:=\left(1+2\frac{\mathcal{X}\mathcal{L}_{,\mathcal{XX}}}{\mathcal{L}_{,\mathcal{X}}}\right)^{-1},
\end{equation}
with is the inverse of the expression showed in~(\ref{hiper}).

Two other hydrodynamic quantities still can be defined, the concentration of particles
\begin{equation}
n\equiv\exp\left[\int\frac{d\rho}{\rho+p(\rho)}\right]=\sqrt{\mathcal{X}}\mathcal{L}_{,\mathcal{X}},
\end{equation}
and enthalpy
\begin{equation}
h:=2\sqrt{\mathcal{X}}.
\end{equation}

We then link the kinetic structure to enthalpy, which is a quantity associated with the capacity of absorption and emission of energy. Enthalpy is a crucial element in theoretical modeling and numerical simulation of astrophysical systems such as neutron stars and black holes, allowing for a deeper understanding of the physical processes that occur in these extreme environments.

\subsection{\emph{K}-essence purely kinetic: Effective Geometric Approach}
In many situations in astrophysics and cosmology, the complete spacetime metrics are extremely complex and difficult to manipulate, which makes it difficult to understand and analyze the physical properties of the system. In such cases, it can be useful to use a simplified space-time description, which can be derived from the full metric, but which takes into account only the properties most relevant to the situation at hand.

The conserved current~(\ref{cc}) allows us to define an effective metric of  $\tilde{\mathcal{G}}^{\mu\nu}$,
\begin{equation}\label{metefe}
\tilde{\mathcal{G}}^{\mu\nu}=\mathcal{L}_{,\mathcal{X}}g^{\mu\nu}+\mathcal{L}_{,\mathcal{X}\mathcal{X}}\nabla_{\mu}\Phi\nabla_{\nu}\Phi.
\end{equation}  

The effetive metric~(\ref{metefe}) has Lorentzian signature and hence describes the time evolution of the system provided and obeys the condition of hyperbolicity~(\ref{hiper}).
If we have $\mathcal{L}_{,\mathcal{XX}}=0$, the~(\ref{metefe}) reduce to $\tilde{\mathcal{G}}^{\mu\nu}=\mathcal{L}_{,\mathcal{X}}g^{\mu\nu}$. Very similar to a conformal transformation, we therefore establish the criteria for the existence of a transformation according to $ \omega^{2} $. It is a scale transformation on the metric of a metric space,   which generates a new metric, associated with another metric space   as $ \mathcal{G}_{\mu \nu} = \omega^{2} {g}_{\mu \nu} $, and which obeys the following requirements:
	\begin{itemize}
    \item $\omega$ has an inverse $ \omega^{-1} $ and is smooth. It has all the derivatives. Transforms the vicinity from a~$ p $ point, in a topological sense, leading to the vicinity of a~$ p'$ point in the transformed metric space. That is loosely equivalent to the properties of an isometry. The expression~(\ref{hiper}) obeys this requeriment; 
    \item The $ \omega $ transformation respects the null geodesic
\begin{equation}\label{geonula}
 g_{\mu\nu}X^{\mu}X^{\nu}=g_{\mu\nu}\omega{x^{\mu}}\omega{x^{\nu}}=\omega^{2}g_{\mu\nu}x^{\mu}x^{\nu}=\mathcal{G}_{\mu\nu}x^{\mu}x^{\nu}=0,
\end{equation}	
which implies that the transformation~$ X^{\mu} = \omega {x^ {\mu}} $ preserves the vector type~(time, space and null). This also implies that $ \omega> 0 $. For simple inspection we notice that~$\mathcal{L}$ and~(\ref{nec}) ensure this requirement;
    \item $ \omega $ respects the angles
		\begin{equation}\label{ang}
			\frac{1}{\sqrt{|v||u|}}g_{\mu\nu}v^{\mu}u^{\nu}=\frac{1}{\sqrt{|V||U|}}\mathcal{G}_{\mu\nu}V^{\mu}U^{\nu}.
		\end{equation}
	The construction of Equation~(\ref{sc}) agrees to the request;
    \item Since $ \omega(a) $ has the $ a $ parameter, it must belong to the original metric space.    
    The~(\ref{sc}) is a scalar product and defines a manifold~$\mathcal{M}$ with metric~$g_{\mu\nu}$.
\end{itemize}.

The criteria established in works~\cite{r1,r2} were respected in this case. A second criterion deducted by Kunhnel and Radmacher~\cite{kunh} has shown a relation between metrics and the Ricci tensor of a space-time, as follows
\begin{equation}\label{ricci1}
\tilde{R}_{ab}-R_{ab}=\frac{1}{\Omega^2}[2\Omega\partial_{\mu}\partial_{\nu}\Omega+(\Omega\partial^{\mu}\partial_{\nu}\Omega-3\partial^{\mu}\Omega\partial_{\mu}\Omega)g_{ab}], 
\end{equation}
where $$\tilde g_{ab}=\omega^2 g_{ab}=\Omega^{-2} g_{ab},$$ so that  $\omega=\Omega^{-1}$. We can see here the difference between the conformal Ricci tensor $\tilde{R}_{\mu\nu}$ and the usual one $R_{\mu\nu}$, related to  $\partial_{\mu}\Omega$, $\partial^{\mu}\partial_{\mu}\Omega$ and $\partial_{\mu}\partial_{\nu}\Omega$. Those operators are Gradient, Laplacian and Hessian of~$ \Omega $ respectively.
The result is that the difference between Ricci tensors after and before the conformal transformation is conserved and proportional to the metric. We write explicitly
\begin{equation}\label{ricci2}
 \tilde{R}_{\mu\nu}-R_{\mu\nu}\propto{g_{\mu\nu}}\propto\tilde g_{\mu\nu}, 
\end{equation}
if, and only if,
\begin{equation}\label{otimo}
 \partial_{\mu}\partial_{\nu}\Omega=(\partial^{\alpha}\partial_{\alpha}\Omega)g_{\mu\nu},  
\end{equation}
where $g_{\mu\nu}$ has zero curvature. 

The Equation~(\ref{otimo}) introduced optimization proprieties for $\mathcal{L}_{\mathcal{X}}$. A very interesting approach can be seen in works~\cite{gwp,k1},  to discuss fluctuations in \emph{k}-essence background where the eikonal is writen as
\begin{equation}
G^{\mu\nu}=\left(\frac{c_{s}}{\mathcal{L}_{,\mathcal{X}}}\right)^{{\frac{1}{\frac{D}{2}-1}}}g^{\mu\nu} .   
\end{equation} 

We consider this approach as a good perspective for future research.

\subsection{The metric of Refractive index perturbation (R.I.P.) frame}
We are going to use works~\cite{rip,rip2} as a guide to write this section.  As a direct application of acoustic formalism, the R.I.P. is a treatment for small fluctuations propagated in a Kerr medium or a dielectric medium. The fluctuations can be characteristic of  locally superluminal propagation and can be described by an eikonal  approximation in a stationary metric.  We introduce the specific analogue model that we have analyzed. As we need to work both in the laboratory frame as in an inertial reference frame, which is moving at relativistic speed relative to the original frame, it is convenient to adopt a covariant formalism. Then consider a reference frame in which the dielectric medium is moving with four-velocity~$u^{\mu}$\cite{rip}. The medium in question can have permittivity and permeability given respectively by
\begin{equation}\label{permi}
\epsilon^{\alpha\beta}:=\epsilon(E)\left(\eta^{\alpha\beta}-u^{\alpha}u^{\beta}\right),
\end{equation}
\begin{equation}\label{permea}
\mu^{\alpha\beta}:=\mu_{0}\mu_{r}\left(\eta^{\alpha\beta}-u^{\alpha}u^{\beta}\right).
\end{equation}

The variable $E:=\sqrt{E_{\mu}E^{\nu}}$ is an electric field, $\mu_{r}$ is the relativity permeability  and~$\mu_{0}$ is the vacuum permeability. Likewise we have 
\begin{equation}
\epsilon(E):=\epsilon_{0}\left(1+\mathcal{X}^{(1)}+\mathcal{X}^{(3)}E^{2}\right).    
\end{equation}

The small fluctuation ``fills'' an effective space-time with two polarization possibilities 
\begin{equation}\label{pol1}
g^{+}_{\mu\nu}=\eta_{\mu\nu}-u_{\mu}u_{\nu}\left(1-\frac{1}{n}\right),
\end{equation}
\begin{equation}\label{pol2}
g^{-}_{\mu\nu}=\eta^{\mu\nu}-u_{\mu}u_{\nu}\left(1-\frac{1}{n^{2}(1+\mathcal{X})}\right) +\frac{\mathcal{X}}{\mathcal{X}+1}l_{\mu}l_{\nu} ,  
\end{equation}
where the vector $l_{\mu}:=\frac{E_{\mu}}{E}$ is unitary in direction of $E_{\mu}$. The parameter $\mathcal{X}$ have an important role since the usual electromagnetism has non generated space-time curvature, because it's a linear field theory. The shift of linearity is measured by parameter $\mathcal{X}$. This parameter is called \emph{nolinearity parameter}
\begin{equation}\label{nled}
\mathcal{X}:=\frac{E}{\epsilon}\frac{\partial\epsilon}{\partial{E}}.    
\end{equation}
Therefore we have a link with non-linear electrodynamics. When $\mathcal{X}\equiv0$, we have linear electrodynamics regime, making the polarization~(\ref{pol1}) and~( \ref{pol2}) match.

After defining those tools, we have to define the refractive index perturbation.
\begin{equation}\label{rip}
n:=n_{0}+\delta{n}(x', \rho) ,   
\end{equation}
and the line-element measured of co-moving frame
\begin{align}\nonumber
ds^{2}:=&c^{2}\frac{\gamma^{2}}{n^{2}}\left(1+\frac{nv}{c}\right)\left(1-\frac{nv}{c}\right)dt'+2\gamma^{2}\frac{v}{n^{2}}\left(1-n^{2}\right)dt'dx' + \\ \label{linrip}
&-\gamma^{2}\left(1+\frac{v}{nc}\right)(dx')^{2}-d\rho^{2}-\rho^{2}d\varphi^{2}. 
\end{align}
The~(\ref{linrip}) have a cylindrical symmetry, $\rho=\sqrt{y^{2}+z^{2}}$. 

Any specific choice for the function $\delta{n}$ give rising to a specific metric in the
aforementioned class. We point out that an isotropic refractive index in the laboratory frame
corresponds to an anisotropic refractive index in the pulse frame due to length contraction
associated with a boost in the x-direction. If the refractive index depends explicitly on~$\rho$, the metric is stationary but not static. It is implied in emergence of the ergoregion, the ergosurface and event-horizon. In this last case, we have a more complex relation, the event-horizon is characterized by  $g_{00}=0$, with $v=\frac{c}{n}$. The event-horizon in these conditions admit   solution in a range 
\begin{equation}\label{rangerip}
\frac{1}{n_{0}+n}<\frac{v}{c}<\frac{1}{n_{0}}.    
\end{equation}

It is necessary highlight a very-important question, in stationary but not static geometry, there is a difference between  event-horizon and ergo-surface, i.e. they do not coincide. The event-horizon in such geometry is in the antipodal points which solve
\begin{equation}\label{antipoda}
    \delta{n}(x', \rho=0):=\frac{c-n_{0}v}{v},
\end{equation}
and ergo-surface is associated with the vanishing of Killing vector~$\partial_{t}$. 

\subsection{Formulation of Einstein-Euller equations}
Following Rezzolla \cite{rezz}, we shall consider a foliation perpendicular to the timeline $\Sigma_{t}$, i.e. a foliation where we can define a perpendicular unitary vector given as follows 
\begin{equation}\label{ee}
\Omega_{\mu}:=\nabla_{\mu}t.
\end{equation}

We then can write the vector defined in~(\ref{ee}) as proportional to another vector 
\begin{equation}\label{ee2}
\eta_{\mu}=A\nabla_{\mu}t, 
\end{equation}
and we can write the tensor $\eta_{\mu}=\left(A,0,0,0\right)$, so we calculate the scalar product $\eta_{\mu}\eta^{\nu}$, where we found 
\begin{equation}\label{norma}
\eta_{\mu}\eta^{\mu}=g^{\mu\nu}\eta_{\mu}\eta_{\nu}=g^{tt}A^{2}.
\end{equation}
The variable $\eta_{\mu}$ corresponds to an observer measuring a four-speed, the timelike condition requires $\eta_{\mu}\eta^{\mu}=-1$, therefore normalization is required. This condition when applied in~(\ref{norma}) gives us
\begin{equation}\label{renorma}
g^{tt}A^{2}=-1, 
\end{equation}
where we chose $\alpha:=-\frac{1}{g^{tt}}$, and for a $\eta^{\mu}$ future we have $A=-\alpha$ . Once the vector $\eta_{\mu}$ is specified, we can define a metric for each hypersurface
\begin{equation}\label{gama}
\gamma_{\mu\nu}=g_{\mu\nu}+\eta_{\mu}\eta_{\nu}
\end{equation}
   and
 \begin{equation}\label{gama1}
\gamma^{\mu\nu}=g^{\mu\nu}+\eta^{\mu}\eta^{\nu}.
\end{equation}

The metric $\gamma_{ij}$ corresponds to the spatial part of $\gamma_{\mu\nu}$, we require that $\gamma^{ik}\gamma_{kj}=\delta^{i}_{j}$ to ensure that $\gamma_{\mu\nu}$ and $\gamma^{\mu\nu}$ correspond to inverse metrics of each other. This construction allows $\eta_{\mu}$ and  $\gamma_{\mu\nu}$ to be two useful tools in the description of any four-vector.

Now we are going to set the projectors 
  \begin{equation}\label{proj1}
  \gamma^{\mu}_{\nu}:=g^{\mu\nu}\gamma_{\alpha\nu}.
  \end{equation}
When we set~(\ref{proj1}), we can resume the definition of~(\ref{gama}) and then we write $g^{\mu}_{\nu}+\mathcal{N}^{\mu}_{\nu}$. A simple inspection ensures that $g^{\mu}_{\nu}+\mathcal{N}^{\mu}_{\nu}=\delta^{\mu}_{\nu}+\mathcal{N}^{\mu}_{\nu}$. Where we set 
  \begin{equation}\label{ene}
\mathcal{N}^{\mu}_{\nu}:=\eta^{\mu}\eta_{\nu},
  \end{equation}
 and the projector acting on~(\ref{ene})
  \begin{equation}\label{putz}
      \gamma^{\alpha}_{\mu}\mathcal{N}^{\mu}_{\nu}=0,
	\end{equation} 
	where a generic four-vector can be written as
  \begin{equation}\label{putz2}
U^{\mu}=\gamma^{\mu}_{\nu}U^{\nu}+\mathcal{N}^{\mu}_{\nu}U^{\nu}.
  \end{equation}

One can easily  notice that $\gamma^{\mu}_{\nu}U^{\nu}=V^{\mu}$ has the contravariant component $V^{t}=0$. On the other hand $V_{t}=g_{t\mu}V^{\mu}\neq0$, as well as the scalar product is given by  $\eta^{\mu}\Omega_{\mu}=\frac{1}{A}\eta^{\mu}\eta_{\mu}=\alpha^{-1}\neq1$, agreeing to the unitary vector defined in~(\ref{ee}). Perpendicular to a hypersurface  spacelike $\Sigma_{t}$, it does not represent changes along the temporal coordinate, and it is not the direction of the time derivative.  The vector that represents the temporal direction is
  \begin{equation}\label{timeita}
\textbf{t}=\textbf{e}_{t}=\alpha\vec{\eta}+\vec{\beta},
  \end{equation} 
	where the vector $\vec{\beta}$ is purely spatial, commonly called the vector shift and with the composition
  \begin{equation}
t^{\mu}\Omega_{\mu}=\alpha\eta^{\mu}\Omega_{\mu}+\beta^{\mu}\Omega_{\mu}=\frac{\alpha}{\alpha}=1.
  \end{equation}
 
The relation above implies the explicit construction of the Eulerian base
  \begin{equation}\label{eu1}
    \eta_{\mu}=\left(-\alpha,0,0,0\right)  
  \end{equation}
  and
  \begin{equation}\label{eu2}
\eta^{\mu}=\frac{1}{\alpha}\left(1,\beta^{i}\right).
  \end{equation}
  
	By Rezzolla \cite{rezz} we write the acoustic line element
  \begin{equation}\label{metiano}
ds^{2}=-\left(\alpha^{2}-\beta_{i}\beta^{i}\right)dt^{2}+2\beta_{i}dx^{i}dt+\gamma_{ij}dx^{i}dx^{j},
  \end{equation}
 which  allows us to identify the vector $\vec{\beta}$ with the velocity of a fluid, calculating  therefore the product to scale with the equations~(\ref{eu1}) and~( \ref{eu2}), 
  \begin{equation}
\eta_{\mu}\eta_{\nu}\eta^{\mu\nu}=\alpha^{2}     
  \end{equation}
  \begin{equation}
  \eta^{\mu}\eta^{\nu}\eta_{\mu\nu}=\frac{1-\vec{\beta}\cdot\vec{\beta}}{\alpha^{2}}.
  \end{equation}
  
	We realize then that $\alpha$ is associated with the spread of signals and $\vec{\beta}$ to a fluid seeping. Finally we write a four-speed vector using the Eulerian elements written above~(\ref{proj1}), (\ref{eu1})
  \begin{equation}
\gamma^{i}_{\mu}u^{\mu}=u^{i},
  \end{equation}
 where we  have the spatial component of a four-speed and 
  \begin{equation}\label{ut}
      -\eta_{\mu}u^{\mu}=\alpha{u}^{t},
  \end{equation}
  the temporal component, wich can be written, in the case of  $\beta_{i}=0$, as 
  \begin{equation}\label{lapso}
      d\tau^{2}=\alpha^{2}dt^{2}.
  \end{equation}
  what allows us us to call the function $\alpha$ time-lapse function~\cite{rezz}.
  
  \section{The acoustic stationary but non static geometry linked to a preferred frame $S_{V}$ }
\subsection{An acoustic geometry}
\begin{figure}
  \centering
\includegraphics[scale=0.3]{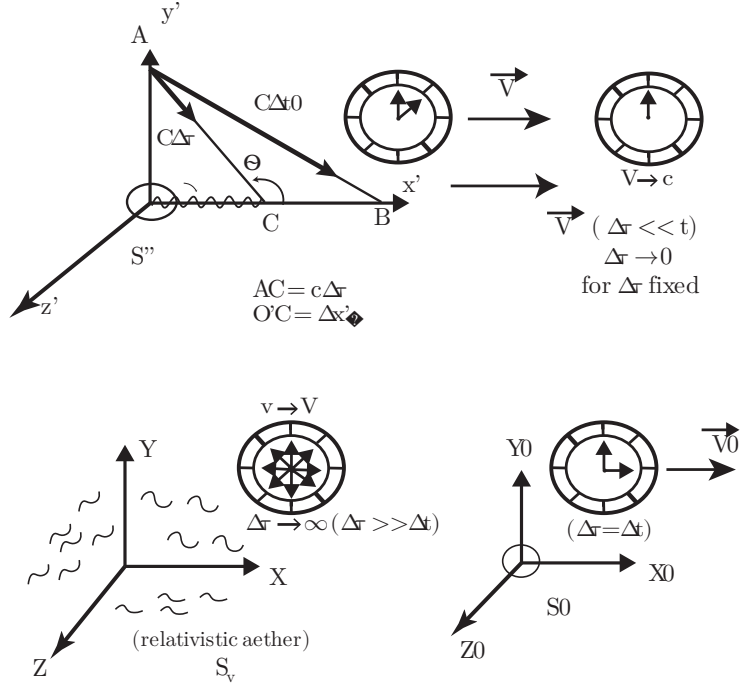}
\caption{The deviation of a photon in reference frames $S'$ and $S_0$ with non null velocities relative to a preferred frame $S_V$, measured by three clocks.}\label{clock}
\end{figure}

Now we are going to analyze the Figure~(\ref{clock}) from a kinematic point of view. It was showed that \cite{N2013}, from a point~\emph{A} located on the axis~$ y'$ of the referential frame~$ S' $, a photon is emitted, in principle, in the direction~$  \overline{AO'} $, as expected for those in the $ S' $ referential frame. However, to our surprise the photon deviates in the direction~$  \overline{AO'} $. This occurs due to the existence of an unreachable minimum speed~$V$, relative to the ultra-reference frame $S_V$. The emitting electron of the photon has an uncertainty in its position in $S'$ due to this minimum velocity, since we cannot say that it is at complete rest. This uncertainty generates a ``delocalization''~$\Delta x'=\overline{O'C}$. Hence, instead of only the segment~$\overline{AO'} $, a rectangle triangle~$\overline{AO'C}$ is formed at the proper frame $S'$ where it is not possible
to find a set of points at rest. For more details of this formalism see also Refs.~{\cite{N2007,N2008,N2015,N2016,N2019,N2022}}.

From the hydrodynamic perspective we also observe that the trajectory (direction) of the excited state seen from~$ S'$ is deviated, because in fact there is no possibility of centering a rigid~$ S '$ coordinate system, whose origin~$ O' $ is fixed precisely on the quasi-particle, since it cannot be located at the origin. Therefore, there is a $ \Delta{x'} = \overline{O'C}> 0 $ ``delocation'' from the origin, and this delocation depends on the speed of~$ S' $ relative to the referential frame $ S_{V} $, i.e, $ \Delta{x'} = \Delta{x'} (v_{S'} / S_{V}) = \Delta{x'(v_{S'})} = \Delta {x'_{v}} $. There is no possibility to undo $ \Delta{x'_{v}} $.
We can show that $ \Delta{x'} \rightarrow \Delta {x'_{min}} $ when $ v\rightarrow{c} $ and  $ \Delta{P} $ increases since $ \Delta{v} \rightarrow{\Delta{v_{max}}} = c $
and that $ \Delta{x_1} \rightarrow \Delta{x'_{max}} $ maximum delocation, when $ v \rightarrow {V} $ ($ \Delta{v} \rightarrow {(\Delta{v}_{min}) = V)}$
in such a way that a certain uncertainty relation $ \Delta{x'_{v}} \Delta{v'} $ is maintained for the particle.
Thus, it is clear that this space-time in the hydrodynamic perspective already has the fundamental ingredients to propose a relation with a  quantum uncertainty. Within an objective framework of reality, now in the Hydrodynamics perspective, essentially quantum.

In Figure~(\ref{clock}), as we have $ \Delta{x'}_{v}> 0 $, we will see two rectangular triangles of a cone, as in relativity. They are:
\begin{enumerate}
    \item $\blacktriangle BO'A $, seen from the external $ S_{0} $ referential frame, which is already separated;
    \item $\blacktriangle BO'C$, obtained in the particle's own $ S'$ referential frame, $  {O'C} $ would be an uncertainty in location (delocation) of it in its own referential $ S'$. As already said, we have $ {O'C} = \Delta{x'}(v) $, where
    
\begin{equation}\label{primeira}
 v = v_{S'/S_{V}}.
\end{equation}
The equation~(\ref{primeira}) represents the speed of a usual reference, measure of the preferred frame $S_{V}$.To be built
\end{enumerate}	

Now, let's extract the following relations from two rectangular triangles:
\begin{enumerate}
    \item $\blacktriangle BO'A: ({AB})^{2}=({O'B})^{2}+({O'A})^2\Rightarrow c^{2}\Delta{t^{2}_{0}}=({O'B})^{2}+v^{2}\Delta{t^{2}_{0}}\Rightarrow$\\
    \begin{equation}\label{31}
({O'B})^{2}=c^{2}\Delta{t^{2}_{0}}-v^{2}\Delta{t^{2}_{0}};
\end{equation}
    \item $\blacktriangle BO'C: ({CB})^{2}=({O'B})^{2}+({O'C})^{2}\Rightarrow c^{2}\Delta{t'}^{2}=({O'B})^{2}+[\Delta{x'(v)}]^{2}\Rightarrow$\\
\begin{equation}\label{32}
{O'B}^{2}=(c\Delta{t'})^{2}-[\Delta{x'_{v}}]^{2}.
\end{equation}
\end{enumerate}

Comparing the relations~(\ref{31}) and~(\ref{32}), follows 
\begin{equation}\label{17}
 c^{2}\Delta{t}^{2}-v^{2}\Delta{t^{2}}=c^{2}\Delta{t'^{2}}-\Delta{x'^{2}_{v}}.
\end{equation} 
where $\Delta\tau=\Delta{t'}, t_{0}=t$, we have
\begin{equation}\label{n1}
 c^{2}\Delta{t}^{2}-v^{2}\Delta{t^{2}}=c^{2}\Delta{\tau^{2}}-\Delta{x'^{2}_{v}}.
\end{equation} 
In Lorentz Space-time, we are (proper) at the origin $ O'$ of $ S' $ so that we have $ \Delta{x'} = 0 $, lose meaning, so we have $ S $ and $ S'$. Therefore, it comes that
\begin{equation}\label{n2}
c^{2}\Delta {t}^{2}-v^{2}\Delta{t}^{2}=c^{2}\Delta{\tau^{2}}=\Delta{S'^{2}}, 
\end{equation}
where $\Delta{\tau}$  is the proper time and $ \Delta{t} $ is the improper one. In this sense, we  define this fundamental  parameter, this speed $v$, expressed in the equation~(\ref{primeira}).
\begin{equation}\label{vparametro}
    v:=\sqrt{c^{2}-\frac{c^{2}(\Delta\tau)^{2}-(\Delta{x}')^{2}}{(\Delta{t})^{2}}}. 
\end{equation}

We take again the~(\ref{n2}) and notice that
\begin{equation}\label{n3}
\left(c^{2}-v^{2}\right)\left(\frac{\Delta{t}}{\Delta\tau}\right)^{2}-c^{2}+\left(\frac{\Delta{x}'}{\Delta\tau}\right)^{2}=0, 
\end{equation}

The equation~(\ref{n3}) is very similar to an acoustic geodesic~\cite{an,enos}
\begin{equation}\label{4tica}
\left(c^{2}_{s}-v^{2}\right)\left(\frac{dt}{ds}\right)^{2}-2v_{i}\left(\frac{dx^{i}}{ds}\right)\left(\frac{dt}{ds}\right)+\left(\frac{dx}{ds}\right)^{2}=0.
\end{equation} 

 When we compare~(\ref{n3}) with~(\ref{4tica}), we get the following coincidence
\begin{equation}\label{rodrigo}
2v\frac{\Delta{x}'}{\Delta\tau}\frac{\Delta{t}}{\Delta\tau}\equiv{c^{2}} .   
\end{equation}
Returning to Equation~(\ref{n3})
\begin{equation}\label{n4}
\left(c^{2}-v^{2}\right)\left(\frac{\Delta{t}}{\Delta\tau}\right)^{2}-2v\frac{\Delta{x}'}{\Delta\tau}\frac{\Delta{t}}{\Delta\tau}+\left(\frac{\Delta{x}'}{\Delta\tau}\right)^{2}=0. 
\end{equation}
The Equations~(\ref{n4}) and~(\ref{4tica}) have  identical formats, then we deduce for relation~(\ref{rodrigo})
\begin{equation}\label{pancadão}
c^{2}=\left(\frac{c^{2}}{v}\frac{\Delta\tau}{\Delta{t}}-v\right)^{2},
\end{equation}
wich can be considered a reissue of the equation~(\ref{vparametro}), but one that allows a better interpretation. Considering the References~\cite{mg,enos},  we interpret the equations~(\ref{n4}) and~(\ref{pancadão}) as the eikonal of a particle,  the speed $v$ is a flux of fluid and $\frac{\Delta{x}}{\Delta\tau}$ as the vector tangent of the geodesic.
Also, the flux and the geometry stay connected.

The next step in procedure is consider a four-vector 
\begin{equation}\label{proibidão}
\frac{\Delta{x'}^{\mu}}{\Delta\tau}=\left(-c\frac{\Delta{t}}{\Delta\tau},\frac{c^{2}}{v}\frac{\Delta\tau}{\Delta{t}}\right).
\end{equation}

Thus we have an apparent violation of Lorentz symmetry, which actually is a deformation of Lorentz invariance, but is recovered by doing~$\Delta{x}=0$.
The equations~(\ref{pancadão}) and~(\ref{proibidão}) indicate a prohibition, the speed~$v$ can not be null. Otherwise we would have a singularity and an inconsistency, the speed $\frac{\Delta{x}}{\Delta\tau}$ may be greater than the light for certain values of $v$, forced the speed to assume a minimum value under certain conditions, this mooring, between the two speeds $\frac{\Delta{x}}{\Delta\tau}$ and $v$ is a real novelty.

To establish a causal relation between the two speeds, it is necessary to resort to the concept of R.I.P.~(\ref{rip}), where the equation~(\ref{antipoda}) represents the relation between event-horizon and R.I.P.

Replacing~(\ref{antipoda}) in~(\ref{rodrigo}), 
\begin{equation}\label{francisco}
\frac{\Delta{x'}}{\Delta\tau}=\frac{c}{2}\frac{\Delta\tau}{\Delta{t}}\left(\eta(x',\rho=0)-n_{0}\right) .   
\end{equation}
Therefore, if the R.I.P. at the origin of the cylindrical coordinate is equal to the refraction index of the medium, we have $\frac{\Delta{x'}}{\Delta\tau}=0$ and we return to an acoustic geometry of Lorentz. This result allows us to interpret the geometric structure we are dealing with as a kind of granular geometric structure, something similar to a vortex because we have cylindrical geometry. So we have more than particles but quasi-particles. We can understand these little fluctuations as phonons. 
A causal link between the two speeds is established, because we associate discrepancy with a quantity that involves a horizon of events. 

We write the new eikonal
\begin{equation}\label{eikonal}
h^{\mu\nu}\frac{{\Delta{x'}_{\mu}}}{\Delta\tau}\frac{{\Delta{x'}_{\nu}}}{\Delta\tau}=0,
\end{equation}

We have an acoustic geometry in particular, associated with a tachionic causal structure.

\subsection{Two-fluid Acoustic Geometry  }
Considering the equations (\ref{4tica}) and (\ref{n3}). We will now subtract  (\ref{4tica}) from (\ref{n3}):
\begin{equation}\label{muitolouca}
\left(1-\frac{v^{2}}{c^{2}}\right)\left(\frac{\Delta{t}}{\Delta\tau}\right)^{2}+1-\frac{1}{c^{2}}\left(\frac{\Delta{x'}}{\Delta\tau}\right)^{2}-\frac{c^{2}_{s}}{c^
{2}}=\frac{1}{c^{2}}||\frac{d\vec{x}}{dt}-\vec{u}||^{2}    
\end{equation}
Here we have that the speed $\vec{v}$ is the velocity of the hypothetical superfluid, measured and compared to the preferred-frame $S_{V}$. The speed $\vec{u}$ is measured in laboratory $\frac{d\vec{x}}{dt}$ is the vector parallel to the flow line, $c_{s}$ is the speed of sound in an acoustic geometry that respects Lorentz symmetry. So the left side contains the sound modified by the presence of the quantities measured by the preferred-frame, the right side the quantities measured in a laboratory. The left side like this, is a sound propagating in this exotic medium. If on the right side we have $\frac{d\vec{x}}{dt}=\vec{u}$, results
\begin{equation}\label{loucalouca}
\left(1-\frac{v^{2}}{c^{2}}\right)\left(\frac{\Delta{t}}{\Delta\tau}\right)^{2}+1-\frac{1}{c^{2}}\left(\frac{\Delta{x'}}{\Delta\tau}\right)^{2}=\frac{c^{2}_{s}}{c^
{2}}   
\end{equation}
the presence of the if the factor , induces sound propagation in a naturally flowing medium , in this case, $c_{s}<c$, if the factor $\frac{\Delta{x'}}{\Delta\tau}=0\rightarrow{c_{s}}=c$. So the waves are electromagnetic. Replacing (\ref{rodrigo}) in (\ref{loucalouca}), we have 
\begin{equation}\label{+doida}
\left( 1-\frac{v^{2}}{c^{2}}\right)\left(\frac{\Delta{t}}{\Delta\tau}\right)^{2}+1-\frac{1}{c}\left(\frac{\Delta\tau}{\Delta{t}}\right)^{2}\left(n(x',\rho=0)-n_{0}\right)^{2}=\frac{c^{2}_{s}}{c^{2}} .
\end{equation}
If $c_{s}=c$, we have $n(x',\rho=0)=n_{0}$, which implies that the medium does not change, we have a continuous medium, with the same refractive index value $n_{0}$. So here we have laminar flow. Therefore, we have the Minkowski space reestablished.
Replacing (\ref{antipoda}) in 
\begin{equation}\label{+doidona}
\left( 1-\frac{v^{2}}{c^{2}}\right)\left(\frac{\Delta{t}}{\Delta\tau}\right)^{2}+1-\frac{c^{2}}{v^{2}}\left(\frac{\Delta\tau}{\Delta{t}}\right)^{2}=\frac{c^{2}_{s}}{c^{2}} .
\end{equation}
The equation (\ref{+doidona}) tells us that values of $v=0$, can be problematic, because we would have a singularity and at the same time a break in causality. We then understand that the terms associated with $\frac{\Delta{x}}{\Delta\tau}$, needs to have material characteristics, and specific material characteristics, which can be related to an index of refraction. Another consequence to be emphasized is that low velocity values imply negative values for the speed of sound, which implies the Jeans' collapse condition \cite{jeans}
\subsection{The Lorentz symmetry violation in an acoustic geometry   }
The tachikonic causal structure usualy correspond to particles more rapid as light. But, we intent to generate a causal structure similar to acoustic tachkion, particles faster than sound. 
In~(\ref{n3})  we have that 

\begin{equation}\label{malfeito}
\Delta{x'}(v) = [w(v)]\Delta\tau. 
\end{equation}
The $ w(v) $ function introduced in Ref.~\cite{N2013} has speed dimension. We interpret~$\Delta{x'}_{v}$ as an internal degree of freedom, then it is better understood as an quasi-particle. 
The function $w(v)$ need some proprieties: 
\begin{enumerate}
\item {dimension of speed};
\item{$w(v)<c$};
\item{$w(v)$ is linked to equation~(\ref{francisco})}.
\end{enumerate}
In physics, the search for fundamental principles is essential, as conservation laws and symmetries. The reciprocity~\cite{valentina,mg,sismo} is a fundamental symmetry for randomic systems and seismology. Surveys in seismology define a quantity called \emph{slowness}~$\left(\frac{1}{v}\right)$, see Ref.~\cite{sismo}, associated to granularity of the medium of propagation p-wave. In quantum turbulence~\cite{superverse} exists a similar conception, the reconnection of two vortexes is associated with speed of flux line $\delta(v)\approx\frac{1}{v}$. But, $w(v)$ have speed dimension, so  a better propouse is   $w(v)=\frac{a}{v}$. In this case $a$ have dimension of quadratic speed, we write $a=v^{2}_{0}$. Furthermore, the $w(v)$ is a opportunity   to introduce reciprocal symmetry
 $w(v)=\frac{v^{2}_{0}}{v}$ to obey the causal structure $v^{2}_{0}=bc$. Where $b$ is a speed, we chose $b=V$ according to Equation~(\ref{n3}), so we have 
 \begin{equation}\label{slow}
w(v)=\frac{cV}{v}.
\end{equation}
We return \ref{slow} in \ref{muitolouca}
\begin{equation}\label{muitolouca1}
\left(1-\frac{v^{2}}{c^{2}}\right)\left(\frac{\Delta{t}}{\Delta\tau}\right)^{2} +1-\frac{V^{2}}{v^{2}}-\frac{c^{2}_{s}}{c^
{2}}=\frac{1}{c^{2}}||\frac{d\vec{x}}{dt}-\vec{u}||^{2}    
\end{equation}
Considering once again $\frac{d\vec{x}}{dt}=\vec{u}$.We must now interpret each of the terms in the acoustic geometry presented, in the equation (\ref{muitolouca1}). 
\begin{enumerate}
\item {The term $\left(1-\frac{v^{2}}{c^{2}}\right)\left(\frac{\Delta{t}}{\Delta\tau}\right)^{2}+1$corresponds to a usual Lorentz factor, including the speed of the wave represented there is the speed of light $c$.}
\item {The term $\frac{c^{2}_{s}}{c^
{2}}=\frac{1}{c^{2}}||\frac{d\vec{x}}{dt}-\vec{u}||^{2}$ corresponds to an acoustic Lorentz symmetry, in the sense proposed by Unhur\cite{unhur} and Visser\cite{vis}. In this geometry we have acoustic tackyons, but, these acoustic tackyons still correspond to time-like shifts for the usual Lorentz geometry. So we have a double cone causal structure. An acoustic cone and a light cone}
\item {The third term  $1-\frac{V^{2}}{v^{2}}$ which was introduced in equation (\ref{slow}). This term is associated with the speed parameter (\ref{vparametro}). This term is associated with the parameter velocity (\ref{vparametro}), so (\ref{vparametro}) is the velocity seen from the referential $S_{V}$, which corresponds to a referential attached to the first excited state of a superfluid, at the moment this excited state appears, in this sense the velocity $v$ is the flow velocity of the hypothetical superfluid A superfluid is characterized by maintaining a flow of energy in the direction of the capillary, but when we deal with the universe there is no capillary, so this velocity corresponds to a rate of departure, between the hypothetical states that would arise, with increasing $v$.We can thus consider the recession velocity of Galaxies deduced by Hubble $v=HX$, where $X$ would be the Galaxies' position \cite{Hubble}. Therefore, when we apply this set of ideas, we imagine that galaxies are analogous to excited states and that the vacuum is a superfluid}
\item{A totally new and different causal structure emerges here, just as particles faster than light describe space-like vectors. Here we have a situation, where two events need to be connected by a vector, which corresponds to a velocity $v>V$, so we have a new range for the validity of the causal structure. For velocities less than $V$, the concept of an event is not defined by the simple laminar flow of this cosmological fluid.  So we write the validity range of the causal structure
\begin{equation}\label{rapina}
V<v<c.
\end{equation} It is necessary to comment on the similarity of this causal structure, with the idea of RIP, described in the equation (\ref{rip}). By pointing out that space-time can go through phase transitions like a fluid, we are not bringing back the idea of a luminiferous aether, but rather conceiving of a model of space-time that would allow us to have different material properties throughout the cosmological ages. Indeed, such a hypothesis is supported by the studies of running constant\cite{rvm}}
\item{ The equation allows us to calculate a new time-lapse function, considering the speed of sound of the acoustic causal structure $c_{s}$ and the granular structure associated with $V$.
\begin{equation}\label{kakaka}
\frac{\Delta{t}}{\Delta\tau}=\frac{1}{c}\sqrt{\frac{c^{2}\left(1-\frac{V^{2}}{v^{2}}\right)-c^{2}_{s}}{1-\frac{v^{2}}{c^{2}}}}
\end{equation}
This result \ref{kakaka} indicates the interaction of three geometries, one the usual geometry of Minkowski space, but compared to the AETHER flow with velocity $v$, a Lorentzian acoustic geometry in the terms the works \cite{unhur} and \cite{vis} represented by $c_{s}$ and the granular term represented by the critical velocity, which is also compared to the AETHER flow with velocity $v$. In the limit where the aether flows with $v=V$, we have a situation where $c^{2}\left(\Delta{t}\right)\equiv-\frac{c^{2}_{s}}{1-\xi^{2}}\left(\Delta\tau\right)^{2}$. It must be said, that the result generalized the works of Unhur \cite{unhur} and Visser \cite{vis}, constituting a more complex acoustic geometry than the acoustic geometry proposed by the \cite{unhur},\cite{vis}. So here we have the Lorentz Symmetry Violation.} 
\end{enumerate}

The introduction of the term 
\begin{equation}\label{Nassif}
   \theta(v):=\sqrt{1-\frac{V^{2}}{v^{2}}}
\end{equation}
modifies the geometry, introducing granularity, being granularity, which is understood as the first excited state, so a very slow particle, a particle that resisted the drag of the cosmic fluid and became very slow, would be dissolved .

When we have $c_{s}=0$  the equation \ref{kakaka} we get the factor 
\begin{equation}\label{newlo1}
\psi(v)=\sqrt{\frac{1-\frac{V^{2}}{v^{2}}}{1-\frac{v^{2}}{c^{2}}}}, 
\end{equation}
$\psi(v)$ is a deformed Lorentz factor, taking the fluid flow as the velocity parameter $v$. So when we eliminate the critical velocity, we get the usual Lorentz factor, see the  \cite{N2007}, \cite{N2008},\cite{N2013},\cite{N2015},\cite{N2016}.
\begin{equation}\label{psi}
\psi(v)=\frac{\Delta{t}}{\Delta\tau}.
\end{equation}

We get:
\begin{enumerate}
    \item $v\rightarrow{c}\Rightarrow\Delta{t_{0}}>>\Delta\tau$(time dilation: the time march in $ S'$ is very slow compared to the time march in $S_{0}$);
    \item $v=v_{0}\Rightarrow\Delta{t_{0}}=\Delta\tau$, here 
    \begin{equation}\label{v0}
        v_{0}=\sqrt{Vc}
    \end{equation}
		\item \textbf{Unprecedented} $v\rightarrow{V}\Rightarrow\Delta\tau>>\Delta{t}_{0}$. Time contraction: the time-march in $ S'$ is much faster compared to the march in $S_{0}$.
\end{enumerate}

\begin{figure}
 \centering
\includegraphics[scale=0.2]{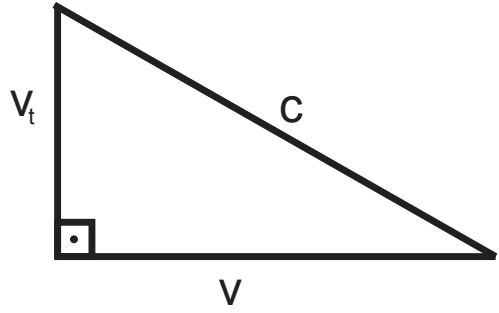}
\caption{ We have $c=\sqrt{v^{2}_{t}+v^{2}}$ -- see~(\ref{muitolouca1}) and~(\ref{timemarch}) -- which represents the
space-temporal speed of any particle (hypothenuse of the triangle$=c$. The novelty here is that such a structure of space-time implements the preferred frame~$S_{V}$ from Ref.~\cite{N2013}}\label{trio}.
\end{figure}

  \subsection{A clock hypothesis associated with the introduction of the privileged reference $S_{V}$}

Once we deduce~(\ref{newlo1}), we can introduce a deformation into Lorentz~(\ref{psi}) and thus, we see the space-time interval considering the existence of the privileged referential of this deformation, we introduce~(\ref{Nassif}) and we rewrite the space-time interval:
\begin{equation}\label{iet}
 c^{2}\Delta\tau^{2}=\frac{1}{1-\frac{V^2}{v^2}}\left[c^{2}\Delta{t}-v^{2}\Delta{t}^{2}\right].
\end{equation}

In Ref.~\cite{N2013,N2019} we see the space-time interval considering the existence of the preferred frame $S_{V}$.

Manipulating~(\ref{iet}) and following~\cite{N2013},  we get 
\begin{equation}\label{timemarch}
 c^2\left(1-\frac{V^2}{v^2}\right)\left(\frac{d\tau}{dt}\right)^{2}+v^{2}=c^{2}. 
\end{equation}
We assume then that the Equation~(\ref{timemarch}) represents the march of time, where the parameter $v$ is the measured speed relative to the preferred frame~$S_{V}$.

 \begin{figure}
  \centering
\includegraphics[scale=0.3]{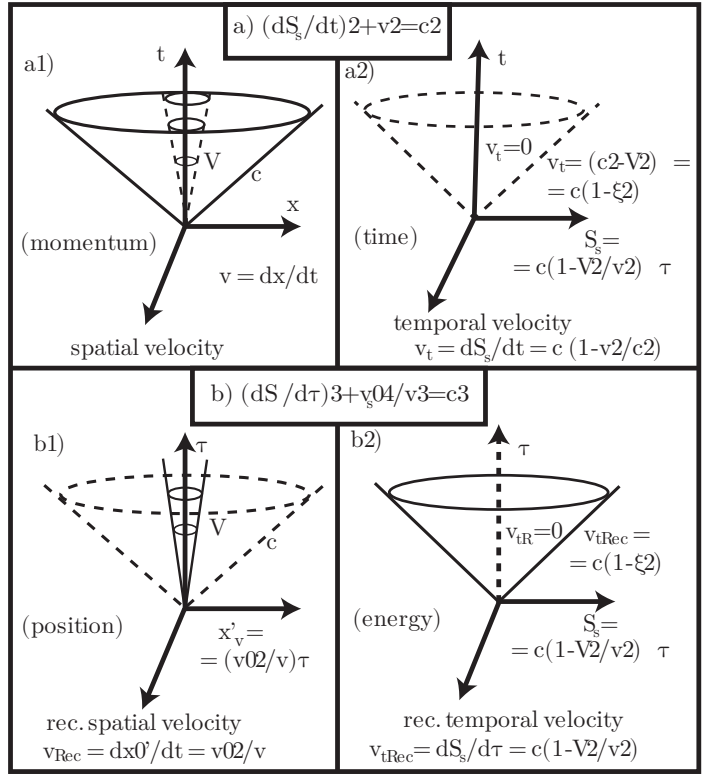}
\caption{ The four cones proposed by Ref.~\cite{N2013} ilustrate an unification of Landau superfluid criticial speed~(\ref{nadau}), the uncertainty-priniple and an acustic causal structure. We see in these figures the causal structures associated with momentum-position and energy-time uncertainty.  }\label{4cones}
\end{figure}

We've rearranged the Equation~(\ref{iet})
  \begin{equation}\label{marin}
   c^{2}\Delta{t}^{2}-v^{2}\Delta{t}^{2}=c^{2}\Delta\tau^{2}-\frac{c^{2}V^{2}}{v^2}\Delta\tau^{2}
  \end{equation}
where we can rewrite the Equation~(\ref{v0}). We can then reverse~(\ref{marin}), writing as
\begin{equation}\label{marchatime}
 c^{2}\left(1-\frac{v^2}{c^2}\right)\left(\frac{dt}{d\tau}\right)^{2}+\frac{v^{4}_{0}}{v^{2}}=c^{2}
\end{equation}

Here we define a speed which we call the \emph{march of space}, as opposed to the march of time $v_{t}$ in Equation~(\ref{marchatime})
\begin{equation}\label{marcha2}
 v_{trec}=c\sqrt{1-\frac{v^{2}}{c^{2}}}\frac{dt}{d\tau}.
\end{equation}

Although the name marches from space may seem unusual, the idea is quite understandable when we think that, in our case, space corresponds to a fluid. We understand that the Equation~(\ref{marcha2}) can be interpreted as the propagation speed of a fluid. We can imagine that~(\ref{marchatime}) is the module of the tangent vector the flow line, parameterized by $\tau$. The Equation~(\ref{marcha2}), would be a complementary fluid to that, analogous to what is done in superfluids. We write then
\begin{equation}\label{timemarch2}
 v^{2}_{trec}+\frac{v^{4}_{0}}{v^2}=c^2.
\end{equation}

Therefore, associating the Equations~(\ref{timemarch}) and~(\ref{marchatime}), we can write a new equation
\begin{equation}\label{recip}
 v^{2}_{trec}+v^{2}_{rec}=c^{2}.
\end{equation}

Now we get a line element, which at the same time is related to $d\tau$ and to $dt$,
\begin{equation}\label{newclock}
 ds_{5}=cd\tau\sqrt{1-\frac{V^2}{v^2}}=cdt\sqrt{1-\frac{v^2}{c^2}}. 
\end{equation}
The Equation~(\ref{newclock})  is a new hypothesis of the deformed clock. Starting from~(\ref{newclock}) we can write 
\begin{equation}\label{linha1}
 \frac{ds_{5}}{d\tau}=c\sqrt{1-\frac{V^{2}}{v^2}},
\end{equation}
and
\begin{equation}\label{linha2}
 \frac{ds_{5}}{dt}=c\sqrt{1-\frac{v^{2}}{c^2}}. 
\end{equation}

The interpretation suggested in our previous work~\cite{nova} for~(\ref{linha1}) and~( \ref{linha2}) discusses the possibility of the two line elements corresponding to a Weyl geometry. And adressing the clock-hypothesis of Weyl for a new perspective \cite{nova}. This kinematics, proposed orginally by Ref.~\cite{N2007} is totally new, because it binds the causal sectors in order to be reciprocal of each other. This reciprocity establishes a structure very similar to wave-particulate duality~(see Figure~\ref{4cones}). 
 
If we do $\frac{ds_{5}}{dt}\frac{d\tau}{ds_{5}}$, we get
\begin{equation}\label{princit}
    \psi=\frac{dt}{d\tau}=\frac{v_{trec}}{v_{t}}\propto{E}.
\end{equation}

In this case $v=\frac{dx}{dt}$ in Equation~(\ref{timemarch}) and~(\ref{slow}) corresponding to an internal structure of the quase-particle. Such internal structure, like the external structure, can be understood as a fluid, also endowed with causal structure as well as the outer fluid. In practice, both external and internal causal structures are associated by~(\ref{linha1}) and~(\ref{linha2}). We call~(\ref{slow}) reciprocal speed.
 Alternatively, if we choose to do $\frac{ds_{5}}{d\tau}\frac{dt}{ds_{5}}$, we get:
\begin{equation}\label{w2}
    \psi^{-1}=\frac{d\tau}{dt}=\frac{v_{t}}{v_{trec}}\propto{time}.
\end{equation}

In this case $v_{t}$ is given for~(\ref{timemarch}), $v_{trec}$ for~(\ref{marcha2}). The Equations~(\ref{princit}) and~(\ref{w2}) indicate the presence of a reciprocal symmetry between the dispersion relation and the principle of energy-time uncertainty. We will study further this issue in the next section.

\subsection{ Principle of Uncertainty in the presence of an hypothetical superfluid }

  The momentum of a massive quasi-particle with respect to
$S_{V}$, with respect to the world line of the particle parameterized by $\tau$ is given by:

\begin{equation}\label{mome}
    P:=m_{0}c\left[\frac{dt}{d\tau}+\Sigma_{i=1}^{3}\frac{dx'_{i}}{d\tau}\right].
\end{equation} 
Where $P$ is the current moment relative to the preferred frame~$S_{V}$. The Equation~(\ref{mome}), when compared to the Equation~(\ref{psi}), suggests a dispersion relation
\begin{equation}\label{maxon}
P^{2}=\frac{m^{2}_{0}v^{2}-m^{2}_{0}V^{2}}{1-\frac{v^{2}}{c^{2}}} ,   
\end{equation}
 very similar to a roton  \cite{superverse}. As~(\ref{princit}), already indicates, the temporal composition of the moment is related to reciprocal symmetry with energy~\cite{mg}, we have a strong indication that we are dealing with excited states of a superfluid. That indication would give a new meaning to the privileged reference frame~$S_{V}$. Therefore, the relation~(\ref{maxon}) is associated with a  differential momentum between the first excited state $P_{V}=m_{0}V$ of a superfluid and the other states $P_{+}=m_{0}v$. This perspective consists of a different interpretation of Nassif's works~\cite{N2008,N2013}. We are considering that the reference~$S_{V}$ is a reference associated with the speed of the first excited state, being dredged by the superfluid. When this reference arises, the fluid itself shifts with minimal velocity, the next states will arise, when the fluid exceeds other speed limits. Thus the difference $\Delta{P}:=\sqrt{P^{2}_{+}-P^{2}_{V}}$ corresponds to the variation of the velocity of the fluid itself in relation to the critical velocity of the fluid, which generates the first excited state. In this perspective, we write:
 \begin{equation}\label{maxonsv}
{P}^{2}=\frac{\Delta{P}^{2}}{1-\frac{v^{2}}{c^{2}}}.
 \end{equation}
 
As example, if $v{\rightarrow}V$ implies $(P\rightarrow0)$ then we have  ${\Delta}x'_{v}\rightarrow\infty$. The new hydrodynamic interpretation makes it possible to understand the delocation of the quasi-particle. When the quasi-particle is slow, it simply dismounts in the superfluid. This leads the understanding of the preferred reference frame~$S_{V}$ more palpable, because actually the reference~$S_{V}$ becomes a conceptual reference frame. It is physical in the meaning of corresponding to a reference with speed where the first excited state arises, then to a virtual observer, which propagates with the emergency speed of the first excited state even if in the system this state is not present.  

  Now we can calculate the following quantity 
\begin{equation}\label{pri}
{\Delta}x'_{v}P=\frac{v^{2}_{0}}{v}{\Delta}t\Psi^{-1}m_{0}v=(m_{0}v_{0})(v_{0}{\Delta}t).
\end{equation}
We can now consider the relation~(\ref{slow}), (\ref{psi}) and~(\ref{mome}). In addotion to $v_{0}=\sqrt{cV}$.
Naturally, we have ${\Delta}x'_{v}P=0$ in special relativity. In Minkowsky space-time $V=0$ so $v_{0}=0$ too. 
  As an analogy, a non-inercial reference frame has the same state of motion given by any acceleration relative to a 
galinean reference frame, thus here we define the flux-line  of the hypothetic superfluid~(\ref{rapina}) relative to a preferred frame
$S_{V}$. 

In other words, we say that every flux-line is a  frame system, these frames systems 
agree with each other as to the given speed of hypothetical superfluid with respect to preferred frame~$S_{V}$. In this meaning, the critical speed of the superfluid is an invariant. Since the speed $V$  is inaccessible for any outher flux-line in a minkowski space-time we have to ${\Delta}v$ is the width of a quasi-particle moving with speed $v$ measurement of a reference associated with a fluid $S_{V}$.

Let's consider a particle that has wavelength $\lambda$ relative to a preferred frame~$S_{V}$ associated with a hypothetical fluid. It would be natural to think that the delocation
$({\Delta}x'_{v})_{0}$, would be given by
\begin{equation}
 \lambda\sim({\Delta}x'_{v})_{0}=\frac{v^{2}}{v_{0}}({\Delta}\tau)_{0}=\frac{v^{2}_{0}}{v}({\Delta}t)_{0}\Psi^{-1},
\end{equation}
where $({\Delta}t)$ will be calculated. We have~(\ref{slow}) using the wavelength definition,
\begin{equation}
\lambda=\frac{h}{P}=\frac{h}{m_{0}v}\sqrt{\frac{1-\frac{v^2}{c^2}}{1-\frac{V^2}{v^2}}}
\end{equation}
and with the Equation~(\ref{mome}) we write: 
\begin{equation}\label{comptoneff}
 \lambda=\frac{h}{m_{0}v}\Psi^{-1}\sim\frac{v^{2}_{0}}{v}({\Delta}t)_{0}\Psi^{-1}.
\end{equation}
From that we get 
\begin{equation}\label{aci}
 m_{0}v^{2}_{0}({\Delta}t)_{0}{\sim}h.
\end{equation}

Finally we compare the equation above~(\ref{aci}) with the Equation~(\ref{pri}) and we have 
\begin{equation}
 ({\Delta}x'_{v})_{0}P=m_{0}v^{2}_{0}({\Delta}t)_{0}{\sim}h.
\end{equation}

Alternatively
\begin{equation}
 ({\Delta}x'_{v})_{0}{\Delta}P=m_{0}v_{0}\lambda_{0}{\sim}h,
\end{equation}
where we have~(\ref{maxonsv}) for any flux-line. The  Equation~(\ref{aci}) is the uncertainty relation of momentum emerging from a space-time with properties of a hypothetically superfluid. 

Now in order to get the relation of uncertainty to energy 
\begin{equation}\label{oi}
 m_{0}cV({\Delta}t)_{0}{\sim}h.
\end{equation}

We can write Equation~(\ref{aci}) as 
\begin{equation}
 m_{0}c^{2}\Psi({\Delta}t)\Psi^{-1}{\sim}h.
\end{equation}

From Equation~$m_{0}c^{2}\Psi({\Delta}t)\Psi^{-1}{\sim}h$ into Equation~(\ref{w2})
\begin{equation}\label{energon1}
 E=m_{0}c^{2}\Psi.
\end{equation}

Thus we write the relation of energy-time uncertainty
\begin{equation}
 E({\Delta}\tau)_{c}=m_{0}c^{2}({\Delta}t){\sim}h,
\end{equation}
or alternatively 
\begin{equation}\label{etq}
 E({\Delta}\tau)_{c}=m_{0}c^{2}\lambda{\sim}h,
\end{equation}
where $\lambda=c({\Delta}t)$. 

We have considered~(\ref{maxonsv}) for any flux line.
We can consider the energy  $E$ relative to a preferred frame~$S_{V}$, we expect uncertainty as far as ${\Delta}E$ for any flux line
$E\equiv{\Delta}E$, in same meaning of~(\ref{maxonsv}). 

  According to the Equation~(\ref{etq}), which is the  energy-time uncertainty principle, we rewrite the Equation~(\ref{w2}) as 
 \begin{equation}\label{4conesb}
\psi^{-1}(v)=\frac{\Delta\tau}{\Delta{t}},
 \end{equation}
 therefore a relation of reciprocity between the clock hypothesis and the principle of energy-time uncertainty is established. The relation of uncertainty becomes an interpretation associated with causality, this causality, which is associated with the emergence of a first excited state, breaks with the laminar flow of a superfluid at the limit where we are close to critical speed. Thus it occurs that  the first excited state breaks the laminar flow, but there is still a drag effect associated with the original laminar flow. We can say
on limit  $v{\rightarrow}V$, it  implies ${\Delta}E\rightarrow0$, where 
\begin{equation}
\psi^{-1}_{v=V}(v)\rightarrow\infty.
\end{equation}

The relathion present on Equation~(\ref{etq}) is an uncertainty time-energy emergent of space-time. In this case the space-time correspond to a hypothetical superfluid~\cite{vortex}.

\subsection{Dispersion Relation in presence of the preferred frame $S_{V}$}
Revisiting the Equations~(\ref{psi}) and~(\ref{w2}), we notice an interesting symmetry
\begin{equation}
\psi^{-1}(v)=\psi(w), 
\end{equation}
where $w(v)$ is given for Equation~(\ref{slow}), it is an indication that~$w(v)$ plays the role of the velocity of counterflow~\cite{vortex}. Considering this perspective the Equation~(\ref{marchatime}) also gains a hydrodynamic interpretation, as a typical superfluid structure, where the superfluid is modeled as two fluids. However, here the two fluids are complementary, in terms of their causal structures.

The appearance of this new structure and its effects on the causal structure have implications, one of these implications concerns the Lorentz symmetry. Revisiting~(\ref{energon1}) we have a relation between energy and causal strucuture, this relation was adressed initially by Equations~(\ref{w2}) and~(\ref{princit}), knowing that~$\Delta{E}\equiv{E}|_{S_{V}}$. Thus we can write
\begin{equation}\label{liv}
\frac{\Delta{t}}{\Delta\tau}=\frac{E|_{S_{V}}-E_{0}}{m_{0}c^{2}}= \frac{E|_{S_{V}}-E_{0}}{E_{0}}.
\end{equation}

The meaning of the Equation~(\ref{liv}) is very important because it is a virtually identical result obtained by Zloshchatiev~\cite{kgz2010}. According to Zloschatiev work~\cite{kgz2010} we have a Lorentz Invariant Violation (LIV), expressed by the line element being multiplied by a function of the energy in Equations~(\ref{linha1}) and~(\ref{linha2}). The difference from the work of the Zloschatiev is he considers~$E_{0}$ as the scale of energy of quantum-gravitational interactions and we consider~$E_{0}$ as the resting energy. It was demonstrated that the introduction of a minimal speed breaks the symmetry of Lorentz~\cite{r1,r3,r2,gbec}. 
However we have not yet established the size of the Lorentz symmetry shift, certainly the energy function~(\ref{energon1}), is a  conformal function  function as already was showed in~\cite{r1,r3}.

Now, we consider the Equations~(\ref{maxon}) and~(\ref{energon1}) and we write
\begin{equation}\label{disperso}
P^{2}-\frac{E^{2}}{c^{2}}=\frac{p^{2}-m^{2}_{0}V^{2}}{1-\frac{v^{2}}{c^{2}}}-m^{2}_{0}c^{2}\frac{1-\frac{V^{2}}{v^{2}}}{1-\frac{v^{2}}{c^{2}}}.
\end{equation}

A transformation in relativistic energy~$E$ and relativsitic momentum~$P$ can be made~$E\rightarrow\hbar\omega$ and~$P\rightarrow\hbar\omega$. We then rewrite~(\ref{disperso})
\begin{equation}\label{media}
P^{2}-\frac{E^{2}}{c^{2}}=\hbar^{2}\kappa^{2}-\hbar^{2}\omega^{2}\left(1+\frac{c^{2}V^{2}}{v^{2}}\right)-\frac{m^{2}_{0}V^{2}}{1-\frac{v^{2}}{c^{2}}},
\end{equation}
where  (\ref{rip}), is an refractive index perturbation~\cite{rip}. 

We call the Equation~(\ref{francisco}) and we relate to~(\ref{rip})
\begin{equation}
\frac{cV}{v}=\frac{c}{2}\frac{\Delta\tau}{\Delta{t}}\left(\eta(x',\rho=0)-n_{0}\right).
\end{equation}
We have the establishment of a lump, a sector where small fluctuations differentiate the fluid from inside the pit, in relation to the outer fluid. 

The term $-\frac{m^{2}_{0}V^{2}}{1-\frac{v^{2}}{c^{2}}}$ is a part of the maxon~(\ref{maxon}), the important to be highleted is that~$\frac{m^{2}_{0}V^{2}}{1-\frac{v^{2}}{c^{2}}}$ can be infinite. Baceti et-al \cite{valentina} survey a dispersion relation. Baceti \emph{et-al}~\cite{valentina} build a kinematic relation which they call an \emph{isotropic Aether-Frame}. We have
\begin{equation}\label{aeb}
    E^{2}-||\vec{p}||^{2}c^{2}=E^{2}_{0}(\tilde{F}),
\end{equation}
here $\tilde{F}$ is an inertial frame. For Bacetti \emph{et-al}  we are working an ``absolute moving'', where right-side of~(\ref{aeb}) is given for~$E^{2}_{0}(v)$ and~$v$ is measured from frame of aether. The energy~$E(v)$ is an usual function~$\eta(V_{aether},V)$. Constituting a minimalist violation of Lorentz symmetry, actually a ``deformation'' of Lorentz symmetry. 

We compare~(\ref{aeb}) and~(\ref{media}) and visualize an equivalence
\begin{equation}\label{olhaeleai}
E^{2}_{0}\equiv\frac{m^{2}_{0}V^{2}}{1-\frac{v^{2}}{c^{2}}}.
\end{equation}

Therefore, we understand the sector associated with~(\ref{olhaeleai}) as a kind of aether, a kind of energy for a privileged reference frame like what we've been calling reference frame~$S_{V}$ in the above sections.

\section{A two-fluid model in presence of preferred frame~$S_{V}$}
\subsection{The Lorentz Invariance Violations and deformed kinematics transformations}
After the study of the previous sections and the deduction of the deformed Lorentz factor~(\ref{psi}), we are allowed to write the kinematic transformations, similar to Lorentz transformations, which relate the references as follows $S'\rightarrow{S}_{V}$:
\begin{equation}\label{9}
dt'=\psi(v)\left[dt+\left(V-v\right)\frac{dx}{c^{2}}\right],
\end{equation}
and
\begin{equation}\label{100}
dx'=\Psi(v)\left[dx+\left(V-v\right)dt\right].
\end{equation}

Being $S'$ any reference and $S_{V}$ a privileged reference proposed by Nassif in a series of works that begin in~\cite{N2007} being the latest~\cite{N2022}, all of them already cited in the above sections. 
So we also have inverse transformations $S_{V}\rightarrow{S}'$
\begin{equation}\label{ex}
dx=\Psi(v)\left[dt'+\left(v-V\right)dt\right],
\end{equation}
\begin{equation}\label{ex1}
dt=\Psi(v)\left[dt'+\left(v-V\right)\frac{dx'}{c^{2}}\right].
\end{equation}

The transformations~(\ref{9}), (\ref{100}), (\ref{ex}) and~(\ref{ex1}) are do not conceive the existence of a perpendicular direction. In order to proceed our study of superfluids, we need to write broader transformations that conceive a larger number of dimensions.  The transformations deduced in~\cite{N2016} satisfy our need.

Let's therefore define the following perpendicular transformations, the one that relates the references~$S'\rightarrow{S_{V}}$
\begin{equation}\label{p}
    dx_{\perp}=\Psi^{-1}(v)dx',
\end{equation}
and the reverse transformation
$S_{V}\rightarrow{S'}$
\begin{equation}\label{pin}
dx'_{\perp}=\Psi(v)Vdt. 
\end{equation}

Once the equations have been written~(\ref{9}), (\ref{100}), (\ref{p}) and~(\ref{pin}), we write the matrices associated with transformations 
\begin{equation}\label{ida}
\left[\begin{matrix}
x'_{\perp}\\
x'_{\parallel}\\
x'_{0}\\
\end{matrix}\right]=M_{3\times3}(S_{V}\rightarrow{S'})\left[\begin{matrix}
x_{\perp}\\
x_{\parallel}\\
x_{0}\\
\end{matrix}\right]^{S_{V}}.     
\end{equation}

We conceive that the relation~(\ref{ida}) can be undone, carrying an array that would exist as 
\begin{equation}\label{volta}
\left[\begin{matrix}
x_{\perp}\\
x_{\parallel}\\
x_{0}\\
\end{matrix}\right]^{S_{V}}=M_{3\times3}(S_{V}\rightarrow{S'})\left[\begin{matrix}
x'_{\perp}\\
x'_{\parallel}\\
x'_{0}\\
\end{matrix}\right] .   
\end{equation}

The matrix of transformations written by Nassif~\cite{N2016} 
\begin{equation}\label{massif}
M_{2\times2}(S_{V}\rightarrow{S}')=\Psi(v)\left[\begin{matrix}
    1 & i\frac{v}{c}\left(1-\frac{V}{v}\right)\\
    -i\frac{v}{c}\left(1-\frac{V}{v}\right) & 1\\
    \end{matrix}\right],
\end{equation}
provides an idea of what the matrix might be $M_{3\times3}(S_{V}\rightarrow{S'})$, as the same algebraic and kinematic properties need to be respected. Our goal here is to include an element perpendicular to translation, then a non-translational term. In this sense, we have no terms outside the diagonal, apart from those present in~(\ref{massif}), because there is no need to represent translations like the terms outside the diagonal $M_{12}=-M_{21}=i\frac{v}{c}\left(1-\frac{V}{v}\right)$. Thus we can build $M_{1\times3}=M_{3\times1}=M_{1\times2}=M_{2\times1}=0$ and we write the matrix
\begin{equation}\label{m33}
M_{3\times3}(S_{V}\rightarrow{S}')=\Psi(v)\left[
\begin{matrix}
1 & 0 & 0 \\
0 & 1 & i\frac{v}{c}\left(1-\frac{V}{v}\right)\\
0 & -i\frac{v}{c}\left(1-\frac{V}{v}\right) & 1
    \end{matrix}\right].
\end{equation}

It is trivial to find the inverse of $M_{3\times3}(S_{V}\rightarrow{S}')$, we write
\begin{equation}\label{m33-1}
M_{3\times3}(S'\rightarrow{S}_{V})=\Psi^{-1}(v)\left[
\begin{matrix}
1 & 0 & 0 \\
0 & \frac{1}{1-\frac{v^{2}}{c^{2}}\left(1-\frac{V}{v}\right)^{2}} & i\frac{\frac{v}{c}\left(1-\frac{V}{v}\right)}{1-\frac{v^{2}}{c^{2}}\left(1-\frac{V}{v}\right)^{2}}\\
0 & -i\frac{\frac{v}{c}\left(1-\frac{V}{v}\right)}{1-\frac{v^{2}}{c^{2}}\left(1-\frac{V}{v}\right)^{2}} & \frac{1}{1-\frac{v^{2}}{c^{2}}\left(1-\frac{V}{v}\right)^{2}} \end{matrix}\right].
\end{equation}

The composition of relations~(\ref{ida}) and~(\ref{volta}) implies that 
\begin{equation}
M_{3\times3}(S_{V}\rightarrow{S}')M_{3\times3}(S'\rightarrow{S}_{V})=\textbf{I}_{3\times3}
\end{equation}
 where $\textbf{I}_{3\times3}$ is the identity matrix. This implies that the crack of matrices 
\begin{equation}\label{trin}
\left[\textbf{I}_{3\times3},M_{3\times3}(S_{V}\rightarrow{S}'),M_{3\times3}(S'\rightarrow{S}_{V})\right],
\end{equation}
make a group, where $\textbf{I}_{3\times3}$ is the neutral element and the two matrices $M_{3\times3}(S_{V}\rightarrow{S}')$ and $M_{3\times3}(S'\rightarrow{S}_{V})$ are inverse to each other. Given the simplicity of this group, it is trivial to note that it is Abelian. However, it is necessary to comment that the frame~$S_{V}$ is unique, this implies that $X'=M_{3\times3}(S_{V}\rightarrow{S'})X_{V}$ is an operation defined in the group, but $M_{3\times3}(S_{V}\rightarrow{S})X'=M_{3\times3}(S_{V}\rightarrow{S})M_{3\times3}(S_{V}\rightarrow{S'})X_{V}$ is not an operation defined in the group. The existence of another equivalent reference would imply the destruction of this structure. The same applies to a transformation that would $S'\rightarrow{S_{V}}\rightarrow{S_{V}}$. The only possible operation in this case is the identity $I_{3\times3}$ that does $S_{V}\rightarrow{S_{V}}$. In addition, the crack of matrices~(\ref{trin}) does not answer the question of transformations between two distinct reference frames~$S_{V}$ and~$S'$.
As Nassif proposed in \cite{N2015} and \cite{N2016} the set of kinematic transformations would not correspond to a group. This conclusion of Nassif is based on Bacetti \emph{et-al} studies~\cite{mg,bacjhep}.

Relations~(\ref{9}), (\ref{100}) and~(\ref{p}) allow us to write compositions of velocities between the references~$S_{V}\rightarrow{S'}$, so we write the parallel component
\begin{equation}\label{vp}
v'_{\parallel}=\frac{v_{\parallel}-u+V}{1+\frac{v_{\parallel}}{c^{2}}(V-u)}.
\end{equation}
From now on we will just write $v_{\parallel}$ as $v$. 

The perpendicular component is 
\begin{equation}\label{vperp}
v'_{\perp}=\frac{V}{1+\frac{v}{c^{2}}(V-u)}.
\end{equation}

Another possible simplification is to adopt~$u=v$, which is equivalent to considering only the two reference frames:~$S'$, the reference frame on the quasi-particle world line; and, the preferred frame,~$S_{V}$. Then we rewrite~(\ref{vp})
\begin{equation}\label{vpu}
 v'_{\parallel}=\frac{V}{1+\frac{(V-v)v}{c^{2}}}   ,
\end{equation}
and~(\ref{vperp})
\begin{equation}\label{vperpu}
v_{\perp}= \frac{V}{1+\frac{(V-v)v}{c^{2}}}.   
\end{equation}

The Equations~(\ref{vpu}) and~(\ref{vperpu}) are the equations associated with the propagation velocity of any excited quasi-particle in the privileged reference frame, and we fall into the first excited state, when we do $v=V$, which means the quasi-particle is being dragged by this exotic fluid. Besides making it clear that this exotic fluid is isotropic, the components of the velocity are identical in all directions, as a wave. A fluctuation that propagates in the fluid, which serves as a privileged reference and that here, already shows  a property of superfluidity, associated with the term 
\begin{equation}
V-v.
\end{equation}

We write the vector $\vec{v}'$, seen from the referential $S'$
\begin{equation}\label{aah}
\vec{v}'=\left[\begin{matrix}
v'_{\parallel}\\
v'_{\perp}\\
\end{matrix}\right] .   
\end{equation}

We can also find parallel and perpendicular velocities when viewed from the point of view of~$S_{V}$, these transformations are important because are based on the reference frame in the superfluid where the currents are constructed~\cite{vortex,superverse,qgv,an}. We write below the speeds seen in the reference~$S_{V}$
\begin{equation}\label{vsv}
v_{\parallel}=\frac{v+u-V}{1+\frac{v}{c^{2}}(u-V)},
\end{equation}
\begin{equation}\label{vsvp}
    v_{\perp}=\frac{\left[1-\frac{v^{2}}{c^{2}}\left(1-\frac{V^{2}}{v^{2}}\right)^{2}\right]V}{1+\frac{v}{c^{2}}(u-V)},
\end{equation}
where
\begin{equation}\label{ex2}
u=v_{S/S_{V}}.
\end{equation}

Now we write the vector 
\begin{equation}
\vec{v}_{S_{V}}=\left[\begin{matrix}
v_{\parallel}\\
v_{\perp}
\end{matrix}\right]_{S_{V}}.    
\end{equation}
\subsection{A concrete example: Vortex}
 The Visser work~\cite{ergovisser} proposes a flow with a vortex profile
\begin{equation}\label{mat}
\vec{v}=\frac{A\hat{r}+B\hat{\theta}}{r}. 
\end{equation}
 We consider~(\ref{mat}) with relation to preferred frame~$S_{V}$ starting from~(\ref{vsv})
 \begin{equation}\label{possivel}
\vec{v}_{\parallel}(S_{V})=\frac{\sqrt{\frac{A^{2}+B^{2}}{r^{2}}}+u-V}{1+\sqrt{A^{2}+B^{2}}\frac{(u-V)}{rc^{2}}}\textbf{e}
 \end{equation}
 where $\textbf{e}$ this related to Equation~(\ref{mat}) and follows $\textbf{e}=\frac{\vec{v}}{||\vec{v}||}$. We write like this, for~(\ref{vsvp})
 \begin{equation}\label{remoto}
v_{\perp}(S_{V})=\left[\frac{1-\frac{A^{2}+B^{2}}{r^{2}c^{2}}\left(1-\frac{V^{2}r^{2}}{A^{2}+B^{e}}\right)}{1+\sqrt{\frac{A^{2}+B^{2}}{r^{2}}}\frac{u-V}{c^{2}}}\right]V\textbf{e}_{\perp}
\end{equation}

When the flux line is given by~(\ref{mat}). The line element associated with~(\ref{mat}) is 
\begin{equation}
ds^{2}=-c^{2}_{som}dt^{2}+\left(dr-\frac{A}{r}dt\right)^{2}+\left(rd\theta-\frac{B}{r}dt\right)^{2}.
\end{equation}

 The limits of~(\ref{possivel}) and~(\ref{remoto}) are 
\begin{equation}
\lim_{r\rightarrow0}v_{\parallel}\rightarrow\frac{c^{2}}{u-V}
\end{equation}
 and 
\begin{equation}
\lim_{r\rightarrow0}v_{\perp}\rightarrow\frac{c^{2}}{u-V}.
\end{equation}

Without incorporating the effects of preferred frame $S_{V}$, to kinematic transformations does not clearly express the kinematic and geometric role of the first excited state of the superfluid. This paper is clear in the Equations~(\ref{possivel}) and~(\ref{remoto}), both have the vortex drag effect by the fluid in laminar flow with minimal speed $V$. The two equations present a second observer $S_{u}$, expressed by speed $u$, for simplicity we will make sure that we have a single state excited in this case $u\equiv{v}$. We rewrite then~(\ref{possivel}) and~(\ref{remoto})
\begin{equation}\label{possivel1}
\vec{v}_{\parallel}(S_{V})=\frac{2\sqrt{\frac{A^{2}+B^{2}}{r^{2}}}-V}{1+\frac{\sqrt{A^{2}+B^{2}}}{rc^{2}}\left(\frac{\sqrt{A^{2}+B^{2}}}{r}-V\right)}\textbf{e}.
 \end{equation}

 We see that in the vicinity of the center of the vortex the speed has an isotropic behavior. The other limit $r\rightarrow\infty$ causes $v_{\parallel}\rightarrow{u-V}$, so the parallel component with the offset would be related to that of the $S_{u}$, already the other component $v_{\perp}\rightarrow\xi^{2}V$. The dimensionless constant $\xi$ is defined and its physical meaning are better explained in Ref.~\cite{gbec} as
 \begin{equation}
     \xi=\frac{V}{c}=\sqrt{\frac{G m_p m_e}{4\pi\epsilon_0}}\frac{q_e}{\hbar c},
 \end{equation}
where $G$ is the gravitational constant, $m_p$ is the proton mass, $m_e$ and $q_e$ are mass and charge of electron. In the same work~\cite{gbec} it is estimated as  $\xi=1.5302\times 10^{-22}$.

 It is remarkable that for small values of $r$, the denominator goes to infinity faster than the numerator, bringing the expression to zero, thus generating a natural cutoff. Such  parameters $A$ and $B$ are responsible for the causal structure, so let's look for a relation to the radius of the ergoregion. Better said, a relation to which distance the effects of Aether drag are felt 
 \begin{equation}\label{vormeu}
     r_{0}=2\frac{\sqrt{A^{2}+B^{2}}}{V},
 \end{equation}
 it is notable that the effect has a much greater scope than the effect predicted by Visser \cite{ergovisser} $2\frac{\sqrt{A^{2}+B^{2}}}{c}$, curiously the ratio between the two effects is of the order of the constant $\xi$. 

\begin{figure}
 \centering
\includegraphics[scale=0.3]{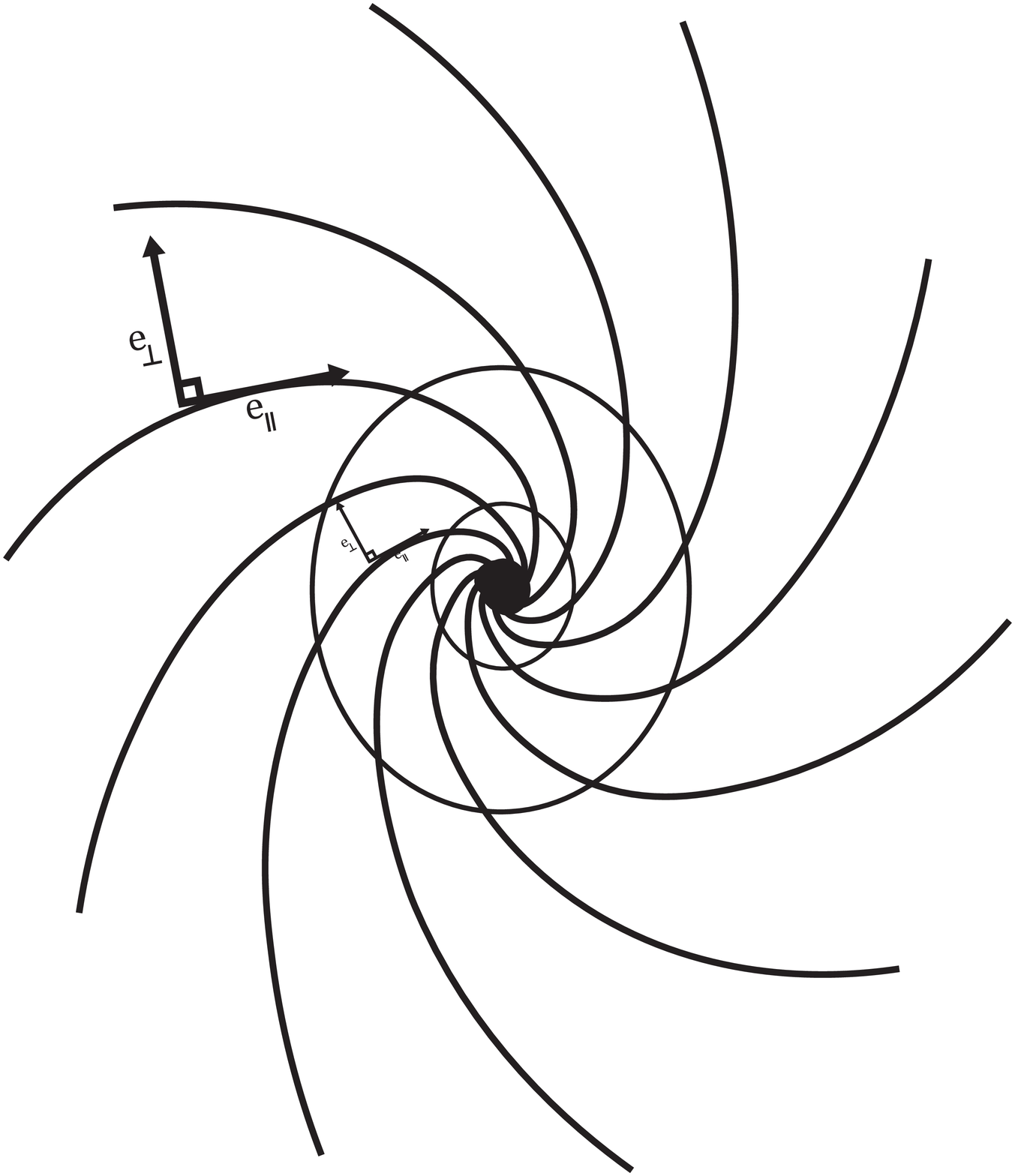}
\caption{ The Vortex Geometry and Ergoregions~\cite{ergovisser} . }\label{fig:vortex}
\end{figure}

 \begin{equation}\label{remoto1}
 v_{\perp}(S_{V})=\left[\frac{1-\frac{A^{2}+B^{2}}{r^{2}c^{2}}\left(1-\frac{V^{2}r^{2}}{A^{2}+B^{e}}\right)}{1+\frac{\sqrt{A^{2}+B^{2}}}{rc^{2}}\left(\frac{\sqrt{A^{2}+B^{2}}}{r}-V\right)}\right]V\textbf{e}_{\perp}
\end{equation}
Similar to~(\ref{possivel1}), we see a cutoff for~$r\rightarrow0$, which indicates that this would be a serene region, like the eye of a hurricane, but the existing velocity would be isotropic, providing a type of expansion, linear velocity in $r$, we then have an effect similar to the Hubble parameter~\cite{vfundo}.  

When we do $\vec{v}_{\parallel}-\vec{v}_{\perp}$ and we consider the limit, we have 
\begin{equation}\label{isotropia}
\lim_{r\rightarrow\infty}||\vec{v}_{\parallel}(S_{V})-\vec{v}_{\perp}(S_{V})||\rightarrow{V},
\end{equation} 
consisting of a region where the vortex influence ceases and we have only the laminar flow of the fluid. 

\subsection{The usual Lorentz symmetry in a superfluid}
We now write transformations of Lorentz, usually for a superfluid 

\begin{equation}\label{3471}
v^{\mu}=(\tilde{\mu},\vec{\nabla}\sigma)=\tilde{\mu}\left(1, \vec{v}_{s}\right),
\end{equation} 
and a perpendicular fluid 
\begin{equation}\label{3472}
v^{\mu}_{n}=\tilde{s}\left(1,\vec{v}_{n}\right).
\end{equation}

We have established a pattern 
\begin{equation}
\tilde{\mu}=\gamma_{s}\mu,
\end{equation}implying in a Lorentz factor
\begin{equation}
\gamma_{s}=\frac{1}{\sqrt{1-\frac{v^{2}_{s}}{c^{2}}}}.
\end{equation}

The same for $\tilde{s}$
\begin{equation}
\tilde{s}=\gamma_{n}s,
\end{equation}
\begin{equation}
\gamma_{n}=\frac{1}{\sqrt{1-\frac{v^{2}_{n}}{c^{2}}}}. 
\end{equation}

The scalar product between~(\ref{3471}) and~(\ref{3472})
\begin{equation}
v^{\mu}_{s}v^{n}_{\mu}=-c^{2}y=-c^{2}\tilde{\mu}\tilde{s}\left(1-\frac{\vec{v}_{s}\cdot\vec{v}_{n}}{c^{2}}\right)=-c^{2}\mu{s}\gamma_{n}\gamma_{s}\left(1-\frac{\vec{v}_{s}\cdot\vec{v}_{n}}{c^{2}}\right), 
\end{equation}
with Lorentz transformations for the addition of $\vec{v}_{s}$ and $\vec{v}_{n}$
\begin{equation}
\vec{v}_{ns}=\frac{\vec{v}_{n}-\vec{v}_{s}}{1-\frac{\vec{v}_{s}\cdot\vec{v}_{n}}{c^{2}}},
\end{equation}

Thus the squared module $||\vec{v}_{ns}||=||\vec{v}_{s}-\vec{v}_{n}||^{2}$ is deliberate
\begin{equation}\label{nortesul}
\frac{v^{2}_{ns}}{c^{2}}=1-\frac{\mu^{2}s^{2}}{y^{4}}.
\end{equation}

Here is established a ratio of the module product of the difference between perpendicular and circulating speeds $||\vec{v}_{s}-\vec{v}_{n}||^{2}$ with the entropy of the system, as a Lorentzian invariant, but there is no lower limit for the speeds of the excited states, there is no information on how excited states behave, in a range where the superfluid begins to generate its first turbulent states. Here is the advantage of the treatment exposed in the Equations~(\ref{vsv}) and~(\ref{vsvp}), since the transformations~(\ref{vsv}) and~(\ref{vsvp}) are more complete than the transformation~(\ref{nortesul}). Although in future work we can integrate the same conceptions set out in~(\ref{nortesul}) with thermodynamic concepts, thus leaving a thermodynamics that takes into account the existence of this superfluid. We also add that the Equation~(\ref{nortesul}) is a referential transformation, a composition of velocities, as well as~(\ref{isotropia}), whereas~(\ref{isotropia}) implies a Lorentz Invariance Violation~(\ref{liv}).

\subsection{Kinematics symmetry compatible with Landau Criteria}
Since we describe the dispersion relation~(\ref{energon1}), we can consider the critical speed generated by it, applying the criterion of Landau~\cite{pai}, \cite{superverse} and~\cite{vortex}. So we rewrite~(\ref{energon1})
\begin{equation}\label{ndefor}
E(v)=m_{0}c^{2}\sqrt{\frac{1-\frac{V^{2}}{v^{2}}}{1-\frac{v^{2}}{c^{2}}}}.
\end{equation}
 Following \cite{vortex}, we consider~(\ref{energon1}), which we rewrite  
 \begin{equation}\label{ndefor1}
E(v)=m_{0}c^{2}\sqrt{\frac{1-\frac{V^{2}}{v^{2}}}{1-\frac{v^{2}}{c^{2}}}}+\vec{p}\cdot\vec{v}_{n}.
\end{equation}

In~(\ref{ndefor1}), we have to~$\vec{v}_{s},\vec{v}_{n}$ are defined as in~(\ref{vsv}) and~$v_{n}$ are defined as in~\cite{vortex} like a perpendicular wind, although here it is more like an induced velocity. The induced expression reflects an analogy with the Faraday-Lens effect. The momentum~$\vec{p}$ is given by the relation~(\ref{maxon}).
We will define another scatter relation 
\begin{equation}\label{correto}
\epsilon(v,p)=m_{0}c^{2}\sqrt{\frac{1-\frac{V^{2}}{v^{2}}}{1-\frac{v^{2}}{c^{2}}}}+\vec{p}\cdot\left(\vec{v}_{n}-\vec{v}_{s}\right).
\end{equation}

Equation~(\ref{ndefor1}) is associated with its excitements.  Thinking of a linear dispersion relation with momentum~$\epsilon(p)\propto{p}$, we consider that the dispersion relation~(\ref{correto}), at the limit where~$p\rightarrow0$ in~(\ref{mome}), we will have~$v\rightarrow{V}$. That implying that, $\vec{v}_{n}+\vec{v}_{s}\rightarrow{V}$, so~(\ref{correto}) will tend to 
\begin{equation}\label{min}
\epsilon(p)\rightarrow{Vp}.
\end{equation} 

That represents the structure of a phonon, as that phonon is associated with a causal structure~(\ref{timemarch}), we can call it acoustic tachyon~\cite{tacos}.
This indicates that, by minimizing~(\ref{correto}), we get 
\begin{equation}\label{nadau}
min\frac{\epsilon(p)}{p}=V,
\end{equation}
and we find a critical speed \cite{superverse,vortex} for the hypothetical gravitational superfluid. Nassif in previous works~\cite{N2015,N2016} had reached a result, associating the limit~(\ref{min}), to the cosmological constant. We find this new kinematic structure, established by Nassif in~\cite{N2008,N2013} and other works, that the cosmologic constant corresponds to the first excited state of a gravitational vacuum. Corresponding directly to Nassif's result in~\cite{N2016}.

\subsection{Conversion between a Eulerian observer and a preferred observer $S_{V}$}
  
   Comparing with the equations of the preferred frame $S_{V}$, the first conclusion we get is 
  \begin{equation}\label{nlapso}
  \alpha\equiv\psi(v),
  \end{equation}
  the second is the immediate comparison between~(\ref{ut}) and~(\ref{timemarch}), that reaffirms~(\ref{nlapso}). Understanding the role of~$\psi(v)$ as a lapse function, we can write the Eulerian base in terms of this new lapse function, via Equations~(\ref{eu1}) and~(\ref{eu2})
  \begin{equation}\label{eunassif1}
    \tilde{\eta}_{\mu}=\sqrt{\frac{1-\frac{V^{2}}{v^{2}}}{1-\frac{v^{2}}{c^{2}}}}\left(-1,0,0,0\right) , 
  \end{equation}
  and
  \begin{equation}\label{eunassif2}
\tilde{\eta}^{\mu}=\sqrt{\frac{1-\frac{v^{2}}{c^{2}}}{1-\frac{V^{2}}{v^{2}}}}\left(1,\beta^{i}\right).
  \end{equation}

It is easy to notice that the scalar product $\tilde{\eta}_{\mu}\tilde{\eta}^{\mu}=-1$ implies that $\tilde{\eta}_{\mu}$ and $\tilde{\eta}^{\mu}$ continue to be unitary, thereby causing 
  \begin{equation}\label{bateu}
g^{tt}=\psi(v).
  \end{equation}
	
  We postulate now.
  \begin{equation}\label{cvd}
  \phi=c^{2}\left(\psi(v)-1\right).
  \end{equation}
  
  Building projectors 
  \begin{equation}\label{gama3}
\tilde{\gamma}_{\mu\nu}=\eta_{\mu\nu}+\tilde{\eta}_{\mu}\tilde{\eta}_{\nu},
\end{equation}
   and
 \begin{equation}\label{gama4}
\tilde{\gamma}^{\mu\nu}=\eta^{\mu\nu}+\tilde{\eta}^{\mu}\tilde{\eta}^{\nu},
\end{equation}
we have established that $\tilde{\gamma}^{ik}\tilde{\gamma}_{kj}$.

We write the new projectors 
\begin{equation}\label{proj2}
  \tilde{\gamma}^{\mu}_{\nu}:=g^{\mu\nu}\tilde{\gamma}_{\alpha\nu}.
  \end{equation}
  
  To determine the four-speed we apply~(\ref{eunassif1}) in~(\ref{ut}) and we have 
  \begin{equation}
    -\tilde{\eta}_{\mu}u^{\mu}=c\sqrt{1-\frac{V^{2}}{v^{2}}}.
  \end{equation}
	
  Applying $\tilde{\eta}^{\mu}$ on both sides 
  \begin{equation}\label{quasela}
    u^{\mu}=-c\sqrt{1-\frac{V^{2}}{v^{2}}}\tilde{\eta}^{\mu},
  \end{equation} 
	remembering that $\tilde{\eta}^{\mu}$ is given in the~(\ref{eunassif2}). 
	
  After we write the Equation~(\ref{quasela}), we can determine by~\cite{rezz}, the speeds $v^{i}:=\frac{\gamma^{i}_{\mu}u^{\mu}}{\alpha{u}^{t}}$ and $v_{i}:=\frac{\gamma_{i\mu}u^{\mu}}{\alpha{u}^{t}}$. We do then 
  \begin{equation}
      \tilde{v}^{i}=-2c\psi^{-2}(v)\beta^{i},
  \end{equation}
  and\begin{equation}
      \tilde{v}_{i}=-2c\eta_{i}=0. 
  \end{equation}
  
	Returning again to Rezzolla \cite{rezz}, using the Equations~(\ref{quasela}) and~(\ref{eunassif2}), we write~$\beta^{i}$, so we get the expression 
  \begin{equation}\label{ptras}
\frac{\beta^{i}}{c}=\left(\psi(v)-1\right)\eta^{i},
  \end{equation}
  bringing back the relation~(\ref{cvd}).         
  
 We also studied the condition of normalization $u^{\mu}u_{\nu}=-1$, to find $u_{\nu}$, we do then $u_{\mu}=a(v)\eta_{\mu}$. Replacing along with~(\ref{quasela}), we found the value of $a=\frac{1}{c\sqrt{1-\frac{V^{2}}{v^{2}}}}$. 
Thus
\begin{equation}
u_{\mu}=\frac{1}{c\sqrt{1-\frac{V^{2}}{v^{2}}}}\tilde{\eta}_{\mu}.
\end{equation}

Studying because 
\begin{equation}\label{k1}
\eta^{\mu\nu}u_{\mu}u_{\nu}=\frac{1}{c^{2}}\frac{1}{1-\frac{V^{2}}{v^{2}}},
\end{equation}
alternatively 
\begin{equation}\label{kesse}
\eta_{\mu\nu}u^{\mu}u^{\nu}=c^{2}\left(1-\frac{V^{2}}{v^{2}}\right).
\end{equation}

 The Equation~(\ref{kesse}) suggest a relation with the \emph{k}-essence models, in the meaning that they are defined in the Equation~(\ref{sc}).

\subsection{Supefluids and \emph{k}-essence  an unified description via preferred frame $S_{V}$}
Considering the work of Nassif et al\cite{r3}. We have the Lagrangian for a relativistic free particle
\begin{equation}\label{50}
\mathcal{L}=-m_{0}c^{2}\sqrt{1-\frac{v^2}{c^2}},
\end{equation} 
however, if the particle suffers effects from a conservative force, which is independent of speed, we have 
$\mathcal{L}=-m_{0}c^{2}\sqrt{1-\beta^{2}}-U$, where $U=U(r)$. 
If the Lagrangian  $\mathcal{L}$  is a function that does not depend on time, if there is no movement constant.

We write the Hamiltonian
\begin{equation}
 h=\dot{q}_{j}p_{j}-\mathcal{L}=\frac{m_{0}v_{j}v_{j}}{\sqrt{1-\frac{v^2}{c^2}}}+m_{0}c^{2}\sqrt{1-\frac{v^2}{c^2}}+U,
\end{equation}
also
\begin{equation}
 h=\frac{m_{0}c^2}{\sqrt{1-\frac{v^2}{c^2}}}+U=E,
\end{equation}
where $h$  is the total energy. 

If  $U=0$, so  $E=\gamma{m_{0}}c^2$, there is a constant movement of the free particle.  We just write 
$v^{2}=v_{j}v^{j}$ for single particle. 

  Now, by using analog procedures to obtain a relativistic Lagrangian, affected by the existence of a preferred frame $S_{V}$ associated with the critical velocity of a superfluid~(\ref{nadau}).
We consider to obtain first Lagrangian of an ideal flow. With $h=E$, where $E$ is get from~(\ref{w2}), the constant movement for a flow-line, knowing the momentum~(\ref{maxon}) is a momentum for the hypothetical superfluid. We write then
\begin{equation}\label{53}
 E=m_{0}c^{2}\sqrt{\frac{1-\frac{V^2}{v^2}}{1-\frac{v^2}{c^2}}}=h=m_{0}v^{2}\sqrt{\frac{1-\frac{V^2}{v^2}}{1-\frac{v^2}{c^2}}}-\mathcal{L},
\end{equation}
where $v^{2}=v_{j}v^{j}$ and  $\mathcal{L}$ is the free particle Lagrangian.

Starting from the Equation~(\ref{53}), we extract:
\begin{equation}\label{54}
 \mathcal{L}=-m_{0}c^{2}\theta\sqrt{1-\frac{v^2}{c^2}}=-m_{0}c^{2}\sqrt{\left(1-\frac{V^2}{v^2}\right)\left(1-\frac{v^2}{c^2}\right)}.
\end{equation}

If we do
 $V\rightarrow0$,  or equivalently~$v>>V$, in Equation~(\ref{54}), we recover the Lagrangian~(\ref{50}). 
 
If we take Equation~(\ref{kesse}) and compare with~(\ref{sc}), we recover the Equation~(\ref{Nassif}). In~(\ref{kesse})  making a change of variable, we get
\begin{equation}\label{kessence}
\mathcal{K}=1-\frac{V^{2}}{v^{2}},
\end{equation}
it is necessary to comment on the range of values that the variable $\mathcal{K}$ can assume
\begin{equation}\label{kin}
0<\mathcal{K}<1-\xi^{2},
\end{equation}

A small manipulation in~(\ref{kessence}) and we have the relation between~(\ref{vparametro}) and~(\ref{kessence})
\begin{equation}\label{prala}
v^{2}=\frac{V^{2}}{1-\mathcal{K}}.
\end{equation}

The Equation~(\ref{prala}) represents the relation between a preferred frame~$S_{V}$ and the \emph{k}-essence typical parameter $\mathcal{K}$, generating an observer coupled to groud-state. It is an imposition for \emph{k}-essence parameter, which is limited to varying within the range, thus constituting a timelike region associated with the causal structure with minimal speed $V$, which simultaneously is the speed criticized according to the criteria of Landau~(\ref{nadau}).

We introduce~(\ref{prala}) in~(\ref{54}) and we obtain 
\begin{equation}\label{k54}
\mathcal{L}(\mathcal{K})=-m_{0}c^{2}\sqrt{\mathcal{K}\left(1-\frac{\xi^{2}}{1-\mathcal{K}}\right)}.
\end{equation}

The Lagrangian~(\ref{k54}) is a \emph{k}-essence purely kinetic Lagrangian just as~(\ref{puro}) and same for the Lagrangian of a superfluid. The consequences and implications  of~(\ref{54}) are implications of generating this unified scenario for cosmology and astrophysics.  The \emph{k}-essence was proposed as a hydrodynamic description for cosmological constant~\cite{k2,kpato}, pathological causal structures in scalar models are mapped~\cite{kpato1} in several works. But~(\ref{k54}) is a different approach. The causal structure of scalar model match with  critical speed, building a preferred frame.   We write  
\begin{equation}\label{lk}
\mathcal{L}_{\mathcal{K}}=-\frac{m_{0}c^{2}}{2\sqrt{\mathcal{K}(1-\frac{\xi^{2}}{1-\mathcal{K}})}}\left(1-\frac{\xi^{2}}{1-\mathcal{K}}-\frac{\xi^{2}\mathcal{K}}{(1-\mathcal{K})^{2}}\right),
\end{equation}
and
\begin{align}\nonumber
\mathcal{L}_{,\mathcal{KK}}=&-\frac{3m_{0}c^{2}}{4\left[\mathcal{K}\left(1-\frac{\xi^{2}}{1-\mathcal{K}}\right)\right]^{\frac{3}{2}}}\left[1-\frac{\xi^{2}}{1-\mathcal{K}}-\frac{\xi^{2}\mathcal{K}}{(1-\mathcal{K})^{2}}\right]+\\\label{lkk}
&-\frac{m_{0}c^{2}}{2\sqrt{\mathcal{K}(1-\frac{\xi^{2}}{1-\mathcal{K}})}}\left(-\frac{\xi^{2}}{(1-\mathcal{K})^{2}}-\frac{1-2\mathcal{K}}{(1-\mathcal{K})^{3}}\right).
\end{align}
The equations permit to write an hydrodynamic approach. The pressure
\begin{equation}\label{pneg}
    \mathcal{P}=\mathcal{L}(\mathcal{K}),
\end{equation}
and density~(\ref{ro}), using~(\ref{kessence}), (\ref{k54}) and~(\ref{lk})
\begin{equation}\label{rok}
\rho=\mathcal{K}\mathcal{L}_{,\mathcal{K}}-\mathcal{L}(\mathcal{K}),
\end{equation}
or even 
\begin{equation}\label{rokp}
\rho=-m_{0}c^{2}\sqrt{\mathcal{K}}\left[\frac{1-\frac{\xi^{2}}{1-\mathcal{K}}\left(1-\frac{\mathcal{K}}{1-\mathcal{K}}\right)}{2\sqrt{1-\frac{\xi^{2}}{1-\mathcal{K}}}}-\sqrt{\left(1-\frac{\xi}{1-\mathcal{K}}\right)}\right].
\end{equation}

Therefore, Equation~(\ref{rokp}) reproduces a vast amount of state equations, with the barotropic parameter
\begin{equation}\label{bp}
w=\frac{\sqrt{\left(1-\frac{\xi}{1-\mathcal{K}}\right)}}{\frac{1-\frac{\xi^{2}}{1-\mathcal{K}}\left(1-\frac{\mathcal{K}}{1-\mathcal{K}}\right)}{2\sqrt{1-\frac{\xi^{2}}{1-\mathcal{K}}}}-\sqrt{\left(1-\frac{\xi}{1-\mathcal{K}}\right)}}.
\end{equation}

The speed of sound is calculated
\begin{align}\nonumber
&c^{2}_{s}=\\\label{barulho}
& \left[1-\frac{\left[-\frac{3}{4}\left[1-\frac{\xi^{2}}{1-\mathcal{K}}\left(1-\frac{\mathcal{K}}{(1-\mathcal{K})^{2}}\right)+\frac{1}{2(1-\mathcal{K})^{2}}\left(\xi^{2}-\frac{1-2\mathcal{K}}{1-\mathcal{K}}\right)\right]\right]\sqrt{\frac{\mathcal{K}}{1-\frac{\xi^{2}}{1-\mathcal{K}}}}}{\frac{1}{2\sqrt{\mathcal{K}\left(1-\frac{\xi^{2}}{1-\mathcal{K}}\right)}}\left[1-\frac{\xi^{2}}{1-\mathcal{K}}\left(1-\frac{\mathcal{K}}{1-\mathcal{K}}\right)\right]}\right]^{-1}.
\end{align}

 The Equations~(\ref{bp}) and~(\ref{barulho}) correspond to the values of the barotropic factor and the sound propagation within the range of values that $\mathcal{K}$ can assume.

\section{A velocity-Potential approach}
Inspired by Rezzola (see details section 3.9 of Ref.~\cite{rezz}), we are going to make a very brief approximation of works~\cite{sch,sch2}. Velocity-potential approach allows the definition of some thermodynamic quantities associated with large kinematics. In this work we will not define the four-speed in terms of the 6 potential scalings of schutz~\cite{sch,sch2} in formalism, because we intend to address the temperature problem associated with event horizons in future work. But the discussion about enthalpy gives back a clarification about the role of the preferred frame~$S_{V}$.

\subsection{The Irrotational perfect fluid}

The hydrodynamic four-speed~(\ref{nor4v})
\begin{equation}\label{forflu}
\mathcal{U}_{\mu}:=\frac{\partial_{\mu}\Phi}{\sqrt{2\left(1-\frac{V^{2}}{v^{2}}\right)}}.    
\end{equation}

The quantity $\mathcal{U}_{\mu}$ represents an irrotational fluid  written in Schutz formalism,~\cite{sch,sch2}, with normalization 
\begin{equation}\label{normalizado}
\mathcal{U}^{\mu}\mathcal{U}_{\mu}=-1,
\end{equation}
which implies
\begin{equation}\label{forfla}
\mathcal{U}^{\mu}:=\sqrt{2\left(1-\frac{V^{2}}{v^{2}}\right)}\partial^{\mu}\Phi. 
\end{equation}

Performing a contraction with Minkowski metric
\begin{equation}\label{schutz}
\eta_{\mu\nu}\mathcal{U}^{\mu}\mathcal{U}^{\nu}=2\left(1-\frac{V^{2}}{v^{2}}\right),
\end{equation}
as in the Schutz formalism\cite{sch}, is given for enthalpy
\begin{equation}\label{entapia}
h(v)=\sqrt{2\left(1-\frac{V^{2}}{v^{2}}\right)},
\end{equation}
a thermodynamic quantity associated the capacity of absorb or emit heat, the simplest substance have enthalpy $h=0$. In this case, the more simple state have $h(v=V)=0$. This results confirm a propriety of ground state, the enthalpy have a relation with pressure  $h=\frac{\rho+P}{n}$, so $h=0$ implies 
\begin{equation}
    P=-\rho.
\end{equation}

We have indication that the cosmological constant is a ground-state of this cosmological superfluid. The density of particle is related with the quantities in this description. The concept of preferred frame~$S_{V}$ then makes it possible to introduce the cosmological constant naturally into a hydrodynamic structure. 
\begin{equation}
n=\sqrt{1-\frac{V^{2}}{v^{2}}}\mathcal{P}_{,\mathcal{K}}    
\end{equation} where $\mathcal{P}_{,\mathcal{K}}=\mathcal{L}_{,\mathcal{K}}$. Using~(\ref{lk}), we rewrite so 
\begin{equation}
    n(\mathcal{K})=-\frac{m_{0}c^{2}}{2\sqrt{(1-\frac{\xi^{2}}{1-\mathcal{K}})}}\left(1-\frac{\xi^{2}}{1-\mathcal{K}}-\frac{\xi^{2}\mathcal{K}}{(1-\mathcal{K})^{2}}\right).
\end{equation}

We have the limits of $n(\mathcal{K}\rightarrow0)\rightarrow{m_{0}}c^{2}$, the $n(\mathcal{K}\rightarrow1-\xi^{2})\rightarrow\infty$. In the first one we recover the classical resting energy, in the second one there is predominance of the \emph{k}-essence and tending to ground-state. The energy-momentum tensor~(\ref{tm}) is written in terms of thermodynamics quantities
\begin{equation}
T_{\mu\nu}=nh\mathcal{U}_{\mu}\mathcal{U}_{\nu}-P(\mathcal{K})\eta_{\mu\nu}    
\end{equation}
Also, we have a hydrodynamic description of \emph{k}-essence superfluid. Thus the Equation~(\ref{forfla}) is an enthalpy current.

\subsection{Conserved current of enthalpy}
We reconsider the Equation~(\ref{forfla}), and we study its conservation
\begin{equation}\label{conserva}
\partial_{\mu}\mathcal{U}^{\mu}=\frac{\sqrt{2}}{\sqrt{1-\frac{V^{2}}{v^{2}}}}\frac{V^{2}}{v^{3}}\partial_{\mu}v\partial^{\mu}\Phi+\sqrt{2\left(1-\frac{V^{2}}{v^{2}}\right)}\partial_{\mu}\partial^{\mu}\Phi
\end{equation}

We consider as Rezzola~\cite{rezz}
\begin{equation}\label{reza}
U=U^{\mu}e_{\mu}
\end{equation}
where $e_{\mu}$ is a vectorial base.
Let us now make a derivation in~(\ref{reza})
\begin{equation}\label{rezabraba}
\partial_{\nu}U=\partial_{\nu}U^{\nu}e_{\mu}+U^{\mu}\partial_{\nu}e_{\mu}
\end{equation}
when we compare~(\ref{rezabraba}) with~(\ref{conserva}). We identify
\begin{equation}
\partial_{\nu}e_{\mu}\mathcal{U}^{\mu}\equiv\sqrt{\frac{2}{1-\frac{V^{2}}{v^{2}}}}\frac{V^{2}}{v^{3}}\partial_{\mu}v\partial^{\mu}\Phi.
\end{equation}

Again by Rezzola \cite{rezz}, we write
\begin{equation}
    \partial_{\nu}e_{\mu}=\Gamma^{\kappa}_{\mu\nu}e_{\kappa}.
\end{equation}

Therefore
\begin{equation}\label{simbol}
\Gamma^{\alpha}_{\alpha\nu}=\sqrt{\frac{2}{1-\frac{V^{2}}{v^{2}}}}\frac{V^{2}}{v^{3}}\partial_{\nu}v,    
\end{equation}
we thus constitute the hydrodynamic equivalent of a connection. The consequences regarding curvature will be investigated in another work, for now the term is immediately identified with some type of shear between flow lines. The appearance of this term in this hypothetical superfluid model is quite interesting, as it allows connections between vortices and intense gravitational fields, a very interesting line of research with many open questions.

When we consider the conservation of current~(\ref{conserva}), we obtain 
\begin{equation}\label{conservacao}
\sqrt{\frac{2}{1-\frac{V^{2}}{v^{2}}}}\frac{V^{2}}{v^{3}}\partial_{\mu}v\partial^{\mu}\Phi+\sqrt{2\left(1-\frac{V^{2}}{v^{2}}\right)}\partial_{\mu}\partial^{\mu}\Phi=0
\end{equation}we see a wave equation for the scalar field, but the wave equation is increased by a term in~(\ref{simbol}), in classical terms, this term can be compared to shear, in quantum terms the reconnection between vortices. The hypothesis of a type of sink or source is also not ruled out and should be investigated in detail in future works.

\section{Conclusions and perspectives}
We revisited the basic of \emph{k}-essence fluid formalism and build a relation of the proposal of a universal minimum velocity under a preferred reference frame $S_V$ with the Landau criterion for superfluids. We did this unification of disjoint concepts to the structure of a \emph{k}-essence fluid. We also identified the enthalpy of a fluid in the Schutz formalism, the function that deforms the Lorentz transforms, introducing the concept of minimum velocity. In this way, we were able to construct an observed \emph{k}-essence fluid from the first excited state of a superfluid. This implies a reinterpretation of the \emph{k}-essence term, thus allowing us to understand a causal structure with the presence of sonic tachyons.

We have found that the hydrodynamic equivalent of a relativistic connection. Such thing could bring light to the discovery of constraints between hydrodynamis of dark superfluids and space-time curvature. The calculation on curvature tensor on a preferred reference frame is something new under the \emph{k}-essence formalism of superfluids.  

We also intend to address the problem of constraining the thermodynamics with the preferred frame and associate its quantities with event horizons in future work. The discussion about enthalpy made here gives back the question about the role of the preferred frame~$S_{V}$.

This theoretical construction allows an association of thermodynamic concepts with causal structures in a very simple way. Using mathematical concepts common in the academic community, such as conformal transformations.

\end{document}